\documentclass[]{aa}

\usepackage{graphicx}
\usepackage{natbib}
\bibpunct{(}{)}{;}{a}{}{,} 

\usepackage{amsmath,amssymb}
\usepackage{url}
\usepackage{fixltx2e}
\usepackage{bm}
\usepackage[normalem]{ulem}

\usepackage{txfonts}
\usepackage{stfloats}
\usepackage{color}

\newcommand{\bq}{\begin{equation}}
\newcommand{\eq}{\end{equation}}

\DeclareMathAlphabet{\mathitbf}{OML}{cmm}{b}{it}

\newcommand{\kms}{\ensuremath{\mathrm{km}\,\mathrm{s}^{-1}}}

\makeatletter

\newcommand{\Rmnum}[1]{\expandafter\@slowromancap\romannumeral #1@}
\makeatother

\newcommand{\hii}{H\,\textsc{ii}}

\begin{document}

\title{Thermal emission from bow shocks I: }
\subtitle{2D Hydrodynamic Models of the Bubble Nebula}
\author{Samuel Green\inst{1,2,3} \and Jonathan Mackey\inst{1,2} \and Thomas J.~Haworth\inst{4} \and Vasilii V.~Gvaramadze\inst{5,6} \and Peter Duffy\inst{3}
        }
\offprints{green@cp.dias.ie}
\institute{
  Dublin Institute for Advanced Studies, Astronomy \& Astrophysics Section, 31 Fitzwilliam Place, Dublin 2, Ireland
  \and
  Centre for AstroParticle Physics and Astrophysics, DIAS Dunsink Observatory, Dunsink Lane, Dublin 15, Ireland
  \and
  School of Physics, University College Dublin, Belfield, Dublin 4, Ireland
  \and
  Astrophysics Group, Blackett Laboratory, Imperial College London, Prince Consort Road, London SW7 2AZ, UK
  \and
  Sternberg Astronomical Institute, Lomonosov Moscow State University, Universitetskij Pr.~13, Moscow 119992, Russia
  \and
  Space Research Institute, Russian Academy of Sciences, Profsoyuznaya 84/32, 117997 Moscow, Russia
}

\date{Draft 13.03.2019 / Received DD Month 2019 / Accepted DD Month 2019}

\abstract{The Bubble Nebula (or NGC\,7635) is a parsec-scale seemingly spherical wind-blown bubble around the relatively unevolved O star BD+60\degr2522. The small dynamical age of the nebula and significant space velocity of the star suggest that the Bubble Nebula might be a bow shock.  We have run 2D hydrodynamic simulations to model the interaction of the central star's wind with the interstellar medium (ISM). The models cover a range of possible ISM number densities of $n=50-200 \, {\rm cm}^{-3}$ and stellar velocities of $v_{\ast}=20-40 \, \kms$. Synthetic H$\alpha$ and 24\,$\mu$m emission maps predict the same apparent spherical bubble shape with quantitative properties similar to observations. The synthetic maps also predict a maximum brightness similar to that from the observations and agree that the maximum brightness is at the apex of the bow shock. The best-matching simulation had $v_{\ast}\approx20 \, \kms$ into an ISM with $n\sim100\, {\rm cm}^{-3}$, at an angle of 60\degr \, with respect to the line of sight. Synthetic maps of soft ($0.3-2$\,keV) and hard ($2-10$\,keV) X-ray emission show that the brightest region is in the wake behind the star and not at the bow shock itself. The unabsorbed soft X-rays have luminosity $\sim10^{32}-10^{33} \, {\rm erg} \, {\rm s}^{-1}$. The hard X-rays are fainter, luminosity $\sim 10^{30} - 10^{31} \, {\rm erg} \, {\rm s}^{-1}$, and may be too faint for current X-ray instruments to successfully observe. Our results imply that the O star creates a bow shock as it moves through the ISM and in turn creates an asymmetric bubble visible at optical and infrared wavelengths, and predicted to be visible in X-rays. The Bubble Nebula does not appear to be unique, it could be just a favourably oriented very dense bow shock. The dense ISM surrounding BD+60\degr2522 and its strong wind suggest that it could be a good candidate for detecting non-thermal emission.}


\maketitle

\section{Introduction}
\label{sec:intro}

Most stars in the universe have winds in the form of gas ejected from their upper atmosphere. The hydrodynamic interaction of such a wind with the surroundings heats the ambient interstellar medium (ISM). For young hot stars with fast winds, a low density bubble is created from this interaction, expanding with time and displacing the ISM. 

\begin{figure*}
	\centering
	\includegraphics[width=0.7\textwidth]{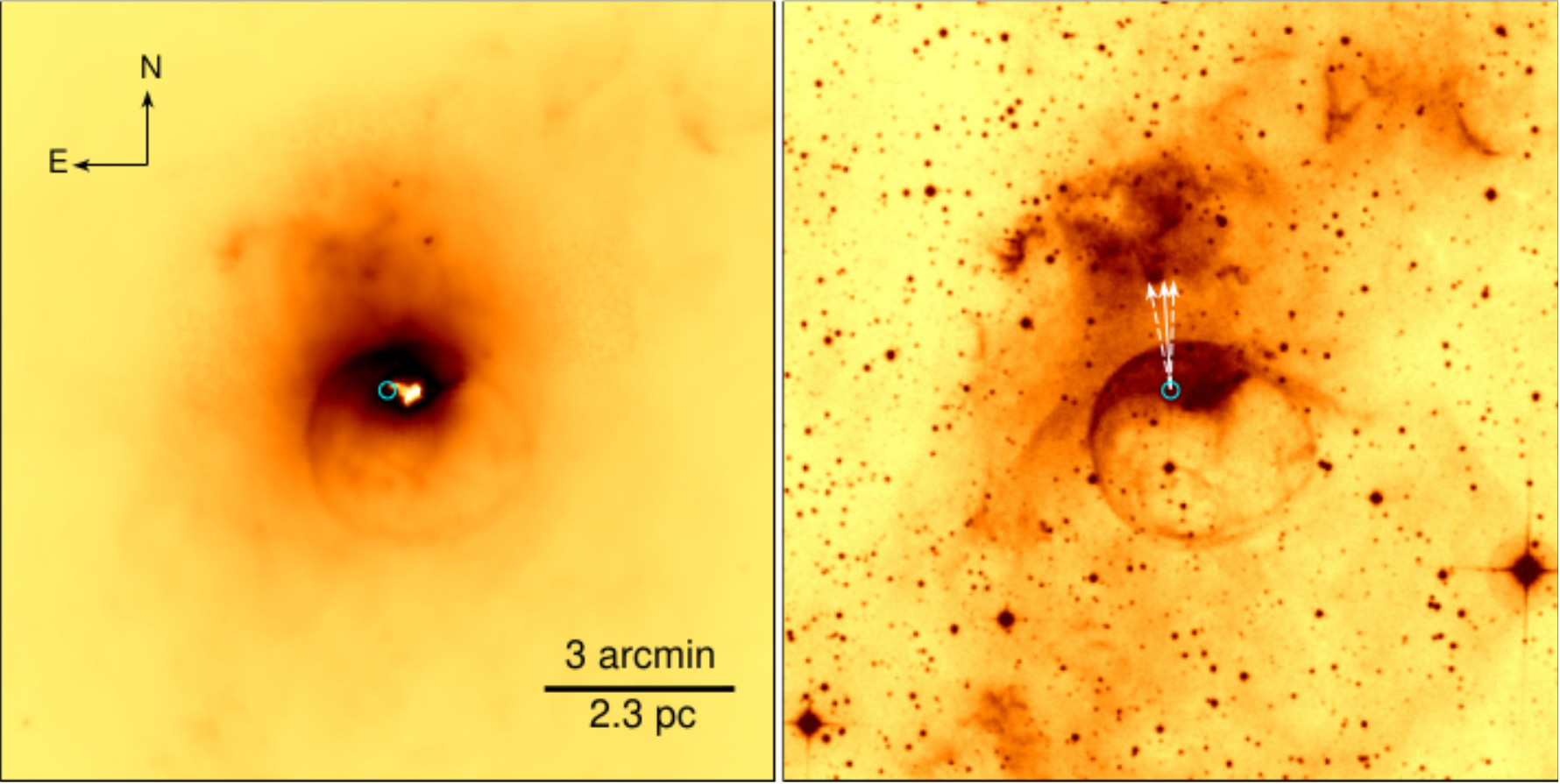}
	\caption{{\it Spitzer} 24\,$\mu$m (left-hand panel) and DSS-II red band (right-hand panel) images of the Bubble Nebula and its surroundings. The arrow shows the direction of motion of BD+60\degr2522 (marked by a circle) as suggested by the {\it Gaia} DR2 proper motion and parallax measurements (with $1\sigma$  uncertainties shown by dashed arrows). The orientation and the scale of the images are the same. A white spot in the {\it Spitzer} image is due to a saturation effect.}
	\label{fig:ngc7635}
\end{figure*}

The physics governing how a wind-blown bubble expands into the ISM is well understood \citep{weaver, 1975ApJ...200L.107C}.
A bubble of hot, shocked gas is formed when a star's wind is converted from kinetic energy to thermal energy as it collides with the surrounding ISM.
Around hot stars, a wind bubble expands within the photoionized H\,\textsc{ii} region that surrounds the star \citep{2006ApJ...638..262F, 2015A&A...573A..10M}, which is usually isothermal with temperature $T\approx10^4$\,K and isothermal sound speed $c_{\rm s}\approx10 \, \kms$.
Young bubbles may expand supersonically into this medium generating a shock wave that sweeps up the ISM into a thin, dense shell, which emits at optical, infrared, and radio wavelengths.

A star in motion with respect to the ISM will generate a bow shock on the upstream side and a turbulent wake downstream.
Up to 25\% of OB stars are indeed moving through space, the so-called runaway stars \citep{Gie87, 1993ASPC...35..207B}, ejected from parent star clusters either because of dynamical encounters with other massive cluster members or due to binary supernova explosions.
Prominent examples are $\zeta$ Oph \citep{GulSof79, 2012MNRAS.427L..50G}, Vela X-1 \citep{KapVanAug97, 2018MNRAS.474.4421G} and BD+43$^\circ$\,3654 \citep{VanNorDga95, 2007A&A...467L..23C}.

Axisymmetric 2D simulations of bow shocks from hot stars were pioneered by \citet{MacVanWoo91} for dense environements and by \citet{1998A&A...338..273C} for the diffuse ISM.
The reverse shock is always adiabatic, as realised by \citet{weaver}, and the forward shock is usually radiative and close to isothermal.
This can lead to the formation of an unstable thin shell.
Thermal conduction may be important at the wind-ISM interface \citep{1998A&A...338..273C, 2014meyer} but may be strongly inhibited by magnetic fields \citep{2017MNRAS.464.3229M}.
Even slowly-moving stars in dense H\,\textsc{ii} regions produce asymmetric wind bubbles \citep{2015A&A...573A..10M} with infrared arcs that look like bow shocks \citep{2016A&A...586A.114M}.
Only stars with very strong winds \citep{2003ApJ...594..888F} and/or moving through dense gas \citep{MacVanWoo91, 2006ApJS..165..283A} drive wind bubbles that fill their H\,\textsc{ii} region, so we are usually justified in modelling bow shocks as photoionized nebulae where hydrogen is fully ionized \citep{2014meyer}, but see \citet{2013MNRAS.431.1337R} for a study in which the surrounding ISM is cold, neutral and turbulent.

Feedback of radiation and energy from massive stars to molecular clouds and the ISM is a key ingredient in understanding the gas dynamics of galaxies \citep[e.g.][]{2015MNRAS.454..238W}.
The important contributions of photoionizing radiation and supernovae is well-established \citep{2004RvMP...76..125M, 2002ApJ...566..302M}, but contributions from stellar winds are much less certain and depend strongly on environment \citep{2018MNRAS.478.4799H}.
If winds are momentum conserving, as argued by \citet{1984ApJ...278L.115M}, then they are relatively unimportant, whereas energy-conserving winds could inject significant kinetic energy and momentum into the ISM \citep{2002ApJ...566..302M}.
X-ray observations are key to solving this issue, because they directly measure the thermal energy content of the bubble.
X-ray observations by \textit{Chandra} and \textit{XMM-Newton} have detected diffuse emission from four Wolf-Rayet bubbles \citep{ChuGruGue03, 2017ApJ...846...76T}, and a number of nascent bubbles and superbubbles around young star clusters \citep{2018ApJS..235...43T}. Early predictions of X-ray luminosities of wind bubbles, however, did not agree with observations quantitatively in that the detected X-ray fluxes were 10 to 100 times lower than those expected \citep{ChuGruGue03}.

As a star moves through the ISM the interaction between its stellar wind and the surrounding medium can produce instabilities that mix material between the adiabatically shocked wind and the photoionized gas in the wake of the bow shock.
This can create a mixing region with plasma temperatures of $\sim 10^6$K \citep{2016ApJ...821...79T}, a strongly cooling boundary layer that efficiently removes thermal pressure support from the bubble.
X-rays originating from this thermal plasma are predicted to exist by several numerical simulations \citep{2006ApJ...638..262F, 2011ApJ...737..100T, 2015A&A...573A..10M, 2014meyer}, at a much lower level than previous predictions \citep[e.g.][]{weaver} because of this wind-ISM mixing.
The existence of this layer also has some support from UV observations \citep{1997ApJ...478..638B}.
On larger scales, observations of X-ray emission from hot gas in star clusters, together with other data, show that the majority of the kinetic energy input by stellar winds is absent and must have escaped the cluster \citep{2011ApJ...731...91L, 2014MNRAS.442.2701R}.

Recently \citet{2016ApJ...821...79T} have detected diffuse emission in the vicinity of the runaway massive star $\zeta$ Oph. They conclude the emission similarly has a thermal nature and its cometary shape agrees with radiation-hydrodynamic models of wind bubbles produced by moving stars. It appears that models and observations are slowly converging, in terms of the X-ray emission, towards the conclusion that stellar wind bubbles are closer to the momentum-conserving limit than energy-conserving. Simulations with different physical assumptions can help us understand these mixing processes, as well as non-thermal processes and stellar wind structures.

In this paper, we are starting a project to investigate thermal emission from stellar wind bubbles. To begin with, we model the Bubble Nebula --- the only known compact (parsec-scale) seemingly spherical nebula around an O star. The main aim of the modelling is to determine whether the shape of this nebula can be explained in terms of the medium around a runaway star. In Sect.\,\ref{sec:obs}, we review observational data on the Bubble Nebula and its associated O star BD+60\degr2522. In Sect.\,\ref{sec:sims}, we describe our model and present the numerical methods and simulation setup. In Sect.\,\ref{sec:sin}, we use our preferred model to produce synthetic H$\alpha$ and infrared emission maps and compare them with observations. In Sect.\,\ref{sec:xray}, we construct synthetic maps of X-ray emission from the model Bubble Nebula and assess the possibility of its detection. We discuss our results in Sect.\,\ref{sec:discusion} and conclude in Sect.\,\ref{sec:conclusions}.

\section{Bubble Nebula and BD+60\degr2522}
\label{sec:obs}

\begin{table}
	\caption{Summary of BD+60\degr2522's parameters. References: (1) \citet{1989ApJS...69..527H}. 
		(2) \citet{1988ApJ...326..356L}. (3) \citet{2018A&A...616A...1G}.}
	\label{tab:param}
	\centering
	\begin{tabular}{l r r} 
		\hline
		Parameter & Value & Refs. \\ [0.5ex] 
		\hline
		Temperature ($T_{\ast}$) & 37\,500 K & (1) \\
		Wind velocity ($v_\infty$) & 2500 km\,s$^{-1} $ & (2) \\
		Mass-loss rate ($\dot{M}$) & $10^{-5.76}$ M$_\odot$ yr$^{-1}$ & (2) \\
		Distance ($d$) & $2.7\pm0.2$ kpc & (3) \\
		Transverse peculiar & & \\
		velocity ($v_{\rm tr}$) & $28\pm3 \, \kms$ & (3) \\ 
		\hline
	\end{tabular}
\end{table}

The Bubble Nebula (or NGC\,7635) is an almost perfectly circular emission nebula of angular diameter of $\approx3$ arcmin. It is clearly visible in infrared and optical wavelengths, as illustrated in Fig.\,\ref{fig:ngc7635} showing the {\it Spitzer Space Telescope} 24\,$\mu$m and the Digitized Sky Survey\,II (DSS-II) red band (McLean et al. 2000) images of the nebula and its surroundings. Fig.\,\ref{fig:ngc7635} also shows that the brightest (northern) side of the Bubble Nebula is faced towards the more extended emission nebula with bright-rimmed clouds to the north, known as SH\,2--162 \citep{1959ApJS....4..257S}. Radial velocity measurements for two nebulae indicate that they are physically associated with each other \citep{1972SvA....16...87D, 1973LIACo..18..357D, 1973A&A....23..147M, 1973A&A....27..143I}

The driving star of the Bubble Nebula is the O6.5\,(n)(f)p \citep{1973AJ.....78.1067W} star BD+60\degr2522. \citet{1971ApJ...170..325C} derived the luminosity class III for BD+60\degr2522, but the peculiar shape of the He\,{\sc ii} $\lambda$4686 emission line in the spectrum of this star makes this assertion uncertain (\citet{2014ApJS..211...10S}; but see below). BD+60\degr2522 is significantly offset from the geometric centre of the nebula towards its brightest edge. Table \ref{tab:param} shows some properties of BD+60\degr2522 including the distance to the star and its peculiar transverse velocity based on the {\it Gaia} second data release (DR2) \citep{2018A&A...616A...1G}. At the distance of $2.7\pm0.2$ kpc, the linear diameter of the Bubble Nebula is $2.3\pm0.2$ pc.

The Bubble Nebula is the only known parsec-scale wind bubble that has been observed around an O star in optical wavelengths. The morphology of the nebula and its neighbourhood has been extensively studied in the 1970s and 1980s where parameters for the central star and the nebula itself were established (see \citet{ChrGouMeaEA95} for a review on the topic). It is generally accepted that the Bubble Nebula is a shell swept up by the stellar wind of BD+60\degr2522 from the dense ($\sim100 \, {\rm cm}^{-3}$) ISM (e.g., \citealt{1973A&A....27..143I, 1986ApJ...306..538V, 1989RMxAA..18...87D, ChrGouMeaEA95, MooWalHesEA02}). The inference on the dense ISM follows from the small linear size of the bubble (e.g. \citealt{ChrGouMeaEA95,MooWalHesEA02}). It is also evidenced by the presence of dense ($\approx10^3-10^4 \, {\rm cm}^{-3}$; e.g. \citealt{2010MNRAS.405.2651M, 2002AJ....124.3305M, 2016MNRAS.460.4038E}) bright-rimmed structures around the Bubble Nebula (one of which is even penetrated into the bubble; e.g. \citealt{2002AJ....124.3305M}), whose ``elephant trunk" morphology is typical of {\hii} regions expanding into dense molecular clouds (e.g. \citealt{1996AJ....111.2349H}). Also, number density estimates based on the [S\,{\sc ii}] $\lambda\lambda$6716, 6731 emission line ratio in the spectrum of the Bubble Nebula showed that the electron number density in its shell is equal to $\approx100-300 \, {\rm cm}^{-3}$ (e.g. \citet{2016MNRAS.460.4038E}; see their Table\,6), which for low-Mach number shocks (see Section\,3.2) corresponds to the pre-shock number density of $\approx50-100 \, {\rm cm}^{-3}$.

The wind blown bubble interpretation of the Bubble Nebula is based on radial velocity measurements \citep{1973LIACo..18..357D, 1983ApJ...274..650L, ChrGouMeaEA95} showing that the central parts of its shell have more positive radial velocities than the rim. This difference in radial velocities implies that we see the far (receding) side of the nebula and that the Bubble Nebula is located on the near side of the molecular cloud associated with SH\,2--162. The illumination of the bright-rimmed clouds surrounding the Bubble Nebula also suggests that BD+60\degr2522 is on the near side of the cloud. If the observed difference in the radial velocities of $\approx20-25 \, \kms$ \citep{ChrGouMeaEA95} represents the expansion of the Bubble Nebula as a whole, then its dynamical age can be estimated to be $5\times10^4$\,yr.

Using $V$ magnitude and $B-V$ colour of BD+60\degr2522 of respectively 8.65 mag and 0.38 mag \citep{1980BICDS..19...61N}, the intrinsic $(B-V)_0$ colour of O6.5 stars of $-0.27$ mag \citep{2005A&A...436.1049M}, and assuming the total-to-selective absorption ratio of $R_V=3.1$, one finds the visual extinction towards the star of $A_V\approx2.0$ mag (which agrees with the extinction estimate based on the Balmer decrement in the spectrum of the Bubble Nebula; see \citet{1972SvA....16..402D}) and its absolute visual magnitude of $M_V=-5.53$ mag. The latter value implies (e.g. \citealt{1973AJ.....78.1067W, 2005A&A...436.1049M}) a luminosity class III for BD+60\degr2522, in agreement with the result by \citet{1971ApJ...170..325C}. The luminosity class III indicates that BD+60\degr2522 is a relatively unevolved star, meaning that the origin of its associated nebula cannot be explained in the framework of the wind-wind interaction scenario proposed for the origin of circumstellar nebulae around evolved massive stars (e.g. \citealt{1996A&A...316..133G, 1996A&A...305..229G}). This inference is supported by chemical abundance measurements for the Bubble Nebula, indicating that it is composed of swept-up ISM \citep{2016MNRAS.460.4038E}.

The brightness asymmetry of the Bubble Nebula and the off-centred location of BD+60\degr2522 could be understood if the nebula impinges on a more dense ambient medium in the north direction (e.g. \citealt{1973A&A....26...45I}) or interacts with a photoevaporation flow from the nearby molecular cloud (e.g. \citealt{MooWalHesEA02}), and/or might be caused by motion of BD+60\degr2522 in the north direction. The latter possibility is supported by the {\it Gaia} DR2 data indicating that BD+60\degr2522 is moving towards the brightest (northern) rim of the Bubble Nebula with a transverse peculiar velocity of $28\pm3 \, \kms$ (see Fig.\,\ref{fig:ngc7635} and Appendix\,A), which is typical of runaway stars.  This in turn suggests that the Bubble Nebula could be a bow shock viewed at an appropriate angle.

To derive the total space velocity of BD+60\degr2522, one needs to know the peculiar radial velocity, $v_{\rm r}$, of this star, which at the distance of 2.7 kpc, is related to the observed heliocentric radial velocity, $v_{\rm r,hel}$, as follows: 
\begin{equation}
v_{\rm r}=v_{\rm r,hel}+39.4 \, \kms \, .
\nonumber
\label{eq:rad}
\end{equation}
The radial velocity of BD+60\degr2522, however, is known to be variable \citep{1952ApJ...115..157W}, which is most likely caused by the line profile variability due to non-radial pulsations typical of the Ofp stars \citep{2003Rauw}. The SIMBAD data base\footnote{http://simbad.u-strasbg.fr/simbad/} provides several values of $v_{\rm r,hel}$ ranging from $-14$ to $-36 \, \kms$. These velocities imply that BD+60\degr2522 is moving either almost in the plane of the sky or receding from us with a velocity comparable to the transverse peculiar velocity. The uncertainty in the stellar velocity relative to the local ISM is aggravated by the possible presence of a photoevaporation flow (whose velocity is of the order of the sound speed, i.e. $\sim10 \, \kms$) from the nearby cloud (cf. \citealt{MooWalHesEA02}) as well as by the radial velocity dispersion of H\,{\sc ii} regions within a spiral arm of $\sim10 \, \kms$ \citep{1976A&A....49...57G}. This means that the total velocity of BD+60\degr2522 relative to the local ISM, $v_\ast$, could range from $\approx20$ to $40 \, \kms$.

Regardless of whether or not a photoevaporative flow is present, the peculiar space velocity of BD+60\degr2522 appears sufficient to create a bow shock. The characteristic scale of the bow shock --- the stand-off distance --- is defined by the balance between the ram pressure of the stellar wind and the ram and thermal pressures of the incoming ISM, and is given by \citep{1970DoSSR.194...41B}:
\begin{equation}
R_{\rm SO}= \sqrt{\frac{\dot{M}v_\infty}{4\pi\rho_\text{ISM}(v^2_{\ast}+c_\text{s}^2)}} \, ,
\label{standoff}
\end{equation}
where $\dot{M}$ and $v_\infty$ are, respectively, the stellar mass-loss rate and wind velocity, and 
$\rho_\text{ISM}$ is the density of the ISM. For bow shocks produced by hot stars (like BD+60\degr2522), $R_{\rm SO}$ gives the minimum distance to the contact discontinuity, separating the shocked stellar wind from the shocked ISM.

In the next sections, we explore the possibility that a bow shock could produce a circular nebula like the Bubble Nebula, which appears to be a closed bubble.

\section{Numerical simulations}
\label{sec:sims}

\subsection{Hydrodynamics and thermodynamics}
We solve the Euler equations of classical hydrodynamics including radiative cooling and heating for optically thin plasma. The equations for the conservation of mass, momentum, and energy are:

\begin{table*}[tp]
	\centering
	\caption{Simulations used for post-processing. Simulations with a `1' in its name have a star velocity of $v_\ast$ = 20 km\,s$^{-1}$, `2' have $v_\ast$ = 30 km\,s$^{-1}$, and `3' have $v_\ast$ = 40 km\,s$^{-1}$. Simulations with an `a' in its name have an ISM ion density of 50 cm$^{-3}$, `b' have  $n_i$ = 100 cm$^{-3}$, and `c' have $n_\text{i}$ = 200 cm$^{-3}$. $v_\ast$ is the star's velocity in km\,s$^{-1}$. $N_{\rm zones}$ shows the number of grid zones in the simulation.}
	\label{tab:sims}
	\begin{tabular}{c c c c c c} 
		\hline
		Name & $v_\ast$ & $n_\text{i}$ &$N_{\rm zones}$ & Box size & Cell size ($\Delta{x}$)\\ [0.5ex] 
		\hline
		1a & 20 & 50 & 1536 $\times$ 1024 & 6.61 $\times$ 4.40 pc & $4.303\times10^{-3}$ pc\\ 
		1b & 20 & 100 & 1536 $\times$ 1024 & 4.67 $\times$ 3.12 pc & $3.040\times10^{-3}$ pc\\
		1c & 20 & 200 & 1536 $\times$ 1024 & 3.30 $\times$ 2.20 pc & $2.148\times10^{-3}$ pc\\
		2a & 30 & 50 & 1536 $\times$ 1024 & 4.40 $\times$ 2.94 pc & $2.864\times10^{-3}$ pc\\
		2b & 30 & 100 & 1536 $\times$ 1024 & 3.12 $\times$ 2.08 pc & $2.031\times10^{-3}$ pc\\
		2c & 30 & 200 & 1536 $\times$ 1024 & 2.20 $\times$ 1.47 pc & $1.432\times10^{-3}$ pc\\
		3a & 40 & 50 & 1536 $\times$ 1024 & 3.30 $\times$ 2.20 pc & $2.148\times10^{-3}$ pc\\
		3b & 40 & 100 & 1536 $\times$ 1024 & 2.34 $\times$ 1.56 pc & $1.523\times10^{-3}$ pc\\
		3c & 40 & 200 & 1536 $\times$ 1024 & 1.65 $\times$ 1.10 pc & $1.074\times10^{-3}$ pc\\
		\hline
	\end{tabular}
\end{table*} 

\begin{align}
\frac{\partial\rho}{\partial{t}} + \nabla \cdot (\rho\bm{v}) &= 0
\label{eqn:cont} \\
\frac{\partial\rho\bm{v}}{\partial{t}} + \nabla \cdot (\bm{v} \otimes \rho\bm{v}) + \nabla{p} &= 0
\label{eqn:momentum} \\
\frac{\partial{E}}{\partial{t}} + \nabla \cdot (E\bm{v}) + \nabla \cdot (p\bm{v}) &= n_\text{e} n_\mathrm{H} \Gamma - n_\text{e} n_\text{i} \Lambda.
\label{eqn:energy}
\end{align}
In equations 
  (\ref{eqn:cont})-(\ref{eqn:energy}),
$\bm{v}$ is the gas velocity in the frame of reference of the star, $\rho$ is the gas mass density,
  $n_\text{i}$, $n_\text{e}$ and $n_\mathrm{H}$ are the number density of ions, electrons and hydrogen nuclei, respectively,
and $p$ is the thermal pressure. $\Lambda$ is the rate for optically-thin radiative cooling and $\Gamma$ is for optically-thin radiative heating. 
$E$ is the total energy density and is its thermal and kinetic parts summed together,
\begin{equation}
E = \frac{p}{(\gamma - 1)} + \frac{\rho{v^2}}{2} \;,
\nonumber
\end{equation}
where $\gamma$ is the ratio of specific heats for a monatomic ideal gas (i.e. $\gamma = 5/3$). The temperature inside a given layer of the bow shock is obtained from the ideal gas law: 
\begin{equation}
T = \mu\frac{m_\text{H}}{k_\text{B}}\frac{p}{\rho} \;.
\nonumber
\end{equation}

The total number density, $n$, is defined by $\rho = \mu n m_\text{H}$, where $\mu$ is the mean mass per particle in units of $m_\text{H}$, the mass of a hydrogen atom.
We consider a gas composed mostly of hydrogen (0.714 by mass), with one helium atom for every 10 of hydrogen, and trace abundances of metals with solar composition.
As discussed above, all of the gas is considered to be photoionized, giving $\mu = 0.61$.
For doubly ionized helium, the electron, ion and hydrogen number densities are given by $n_\text{e} = 0.86  \rho/m_\text{p}$, $n_\text{i} = 0.79 \rho/m_\text{p}$ and $n_\mathrm{H}=0.71 \rho/m_\text{p}$.

The radiative heating is assumed to arise primarily from photoionization of hydrogen atoms that recombine in the H\,\textsc{ii} region and so is simply the product of the recombination rate and a mean heating energy per ionization, $\langle E_\mathrm{pi}\rangle$, \citep[cf.][]{2014meyer}:
\begin{equation}
n_\text{e}n_\text{H} \Gamma = \alpha_\text{B} n_\text{e}n_\text{H} \langle E_\mathrm{pi}\rangle \;.
\nonumber
\end{equation}
We take $\langle E_\mathrm{pi}\rangle=5$\,eV, appropriate for an O star, and use the case B recombination rate $\alpha_\text{B}$ from \citet{Hum94}.
The radiative cooling rate $\Lambda$ includes:
\begin{enumerate}
\item
Metal-line cooling taking the minimum of the cooling curve of \citet{WieSchSmi09} (collisional ionization equilibrium (CIE), metals only) and the forbidden-line cooling function of \citet{HenArtDeC09} (eq. A9, damped exponentially for $T>10^5$\,K).
This captures cooling of shocked wind assuming CIE, and also the strong forbidden-line cooling of the photoionized ISM that would not arise in CIE.
\item
Bremsstrahlung from ionized hydrogen \citep{Hum94} and helium \citep{RybLig79}.
\item Recombination cooling of H, with rate from \citet{Hum94}.
\end{enumerate}

\subsection{Computational methods and initial conditions}

We use the {\sc pion} radiation hydrodynamics code \citep{2012accuracy} to model the Bubble Nebula as a propagating O star emitting a stellar wind. The code solves the Euler equations (Eqs.\,(\ref{eqn:cont})--(\ref{eqn:energy})) in cylindrical coordinates with rotational symmetry on a computational grid in the $(R,z)$ plane. The mass, energy, and momentum densities are defined at the centre of each computational cell, and evolved with time according to the hydrodynamical equations. For a detailed description and explanation of the {\sc pion} code see \citet{2012accuracy}, and for applications of the cylindrical coordinate system see \citet{2012betelgeuse} and \citet{2015A&A...573A..10M}. The integration scheme follows \citet{1991MNRAS.250..581F} and \citet{1998MNRAS.297..265F}.

For numerical convenience, a reference frame in which the star is stationary and located at the origin $(R,z)=(0,0)$ of a rectangular box is chosen. The ISM flows past the star in the negative $z$-direction, interacting with the stellar wind as it does so. A passive scalar variable is utilised to distinguish between the ISM and wind gas. For the sake of simplicity, the ISM is assumed to be homogeneous. 

A range of ISM densities, $n_\text{H}$ = 50, 100, and 200 cm$^{-3}$, with corresponding stellar velocities $v_\ast$ = 20, 30, and $40 \, \kms$, were modelled (see Table \ref{tab:param}). These stellar velocities were chosen to account for uncertainty in the relative velocity of the star (see Sect.\,\ref{sec:obs}). The ISM densities were chosen because this is the range of observed densities derived from the nebular emission lines (see Section 2). The ISM densities and stellar velocities were used to calculate the stand-off distance of the bow shock for each simulation (Eq.\,(\ref{standoff})). This was then used to estimate the size of the simulation boxes.

\begin{figure*}	
	\centering
	\includegraphics[height=.410\textwidth]{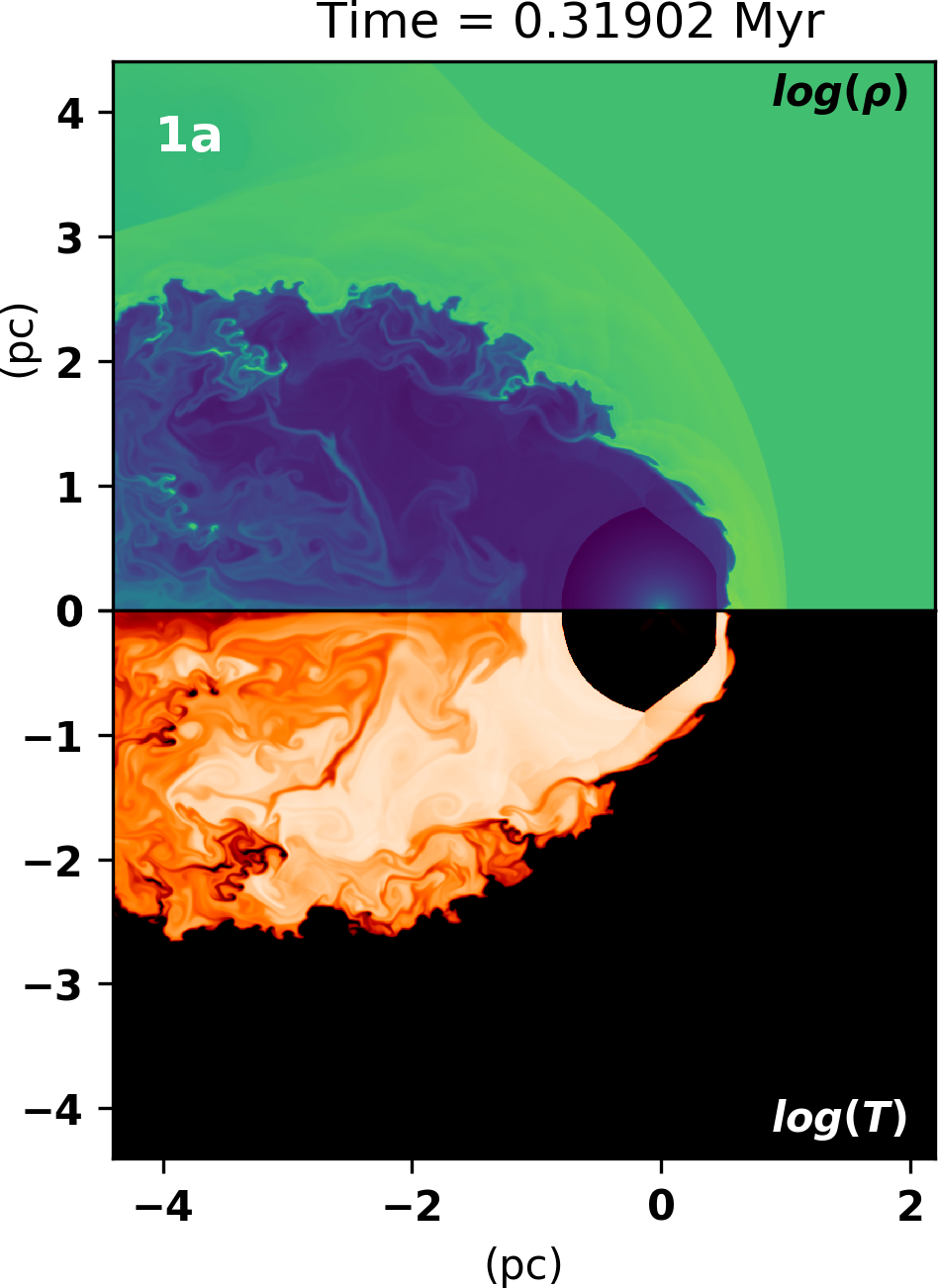}
	\includegraphics[height=.410\textwidth]{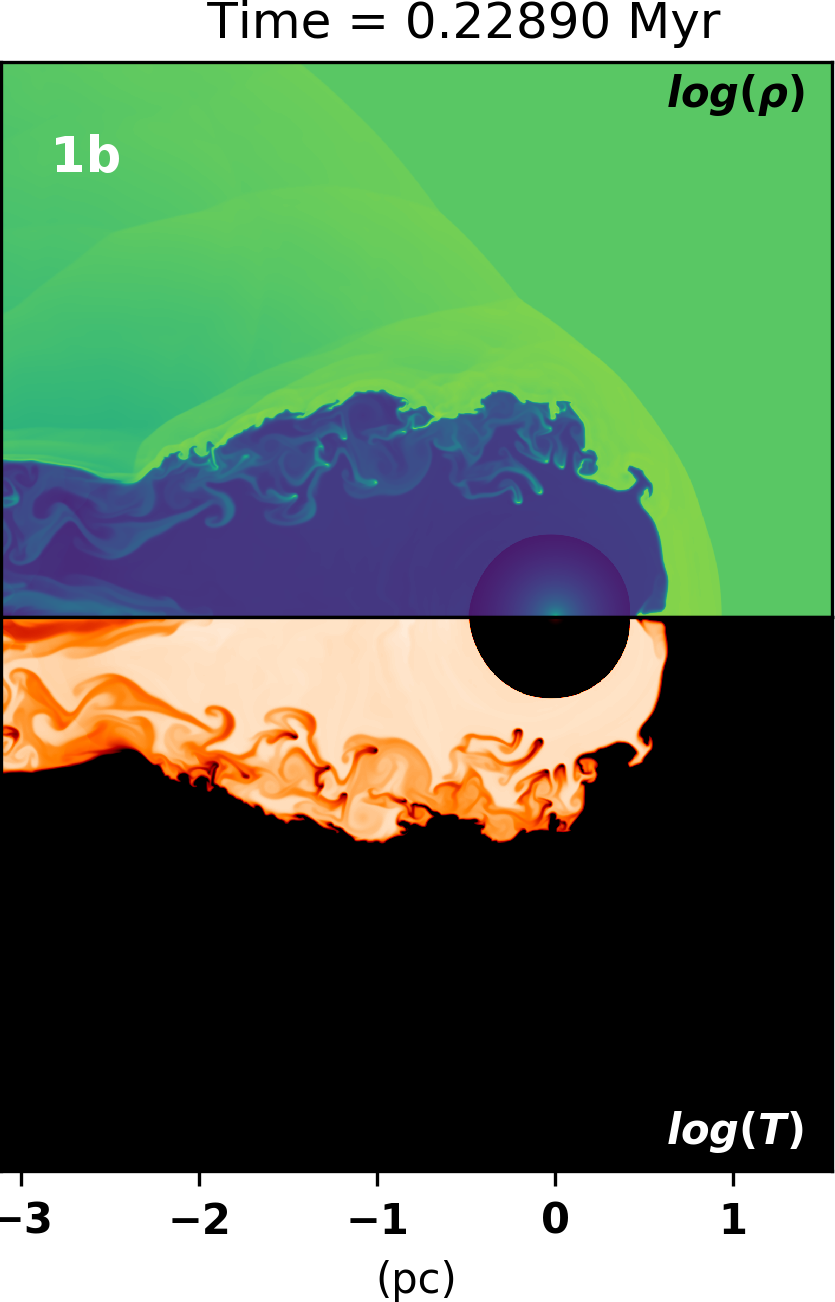}
	\includegraphics[height=.410\textwidth]{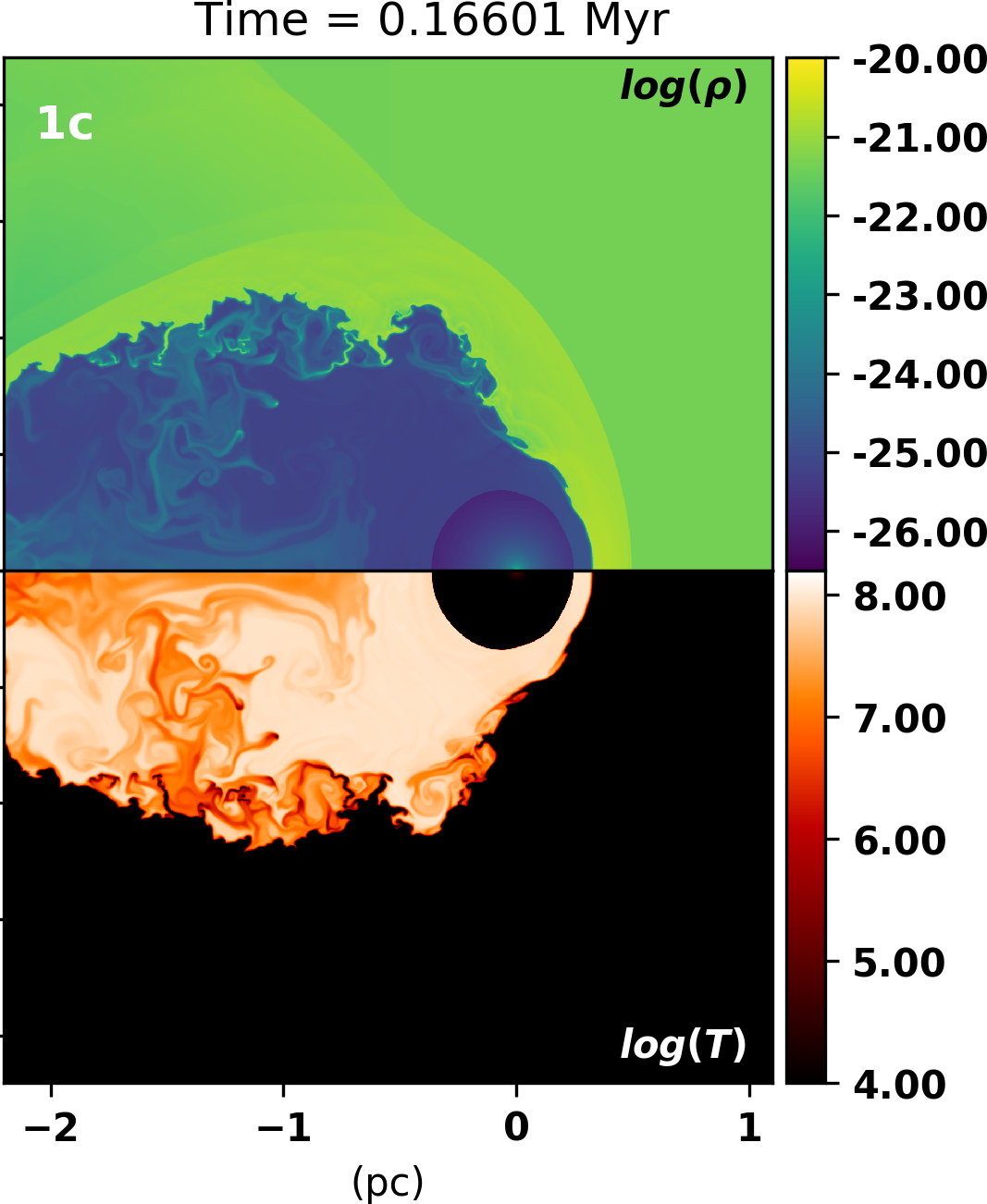} \\
	\includegraphics[height=.410\textwidth]{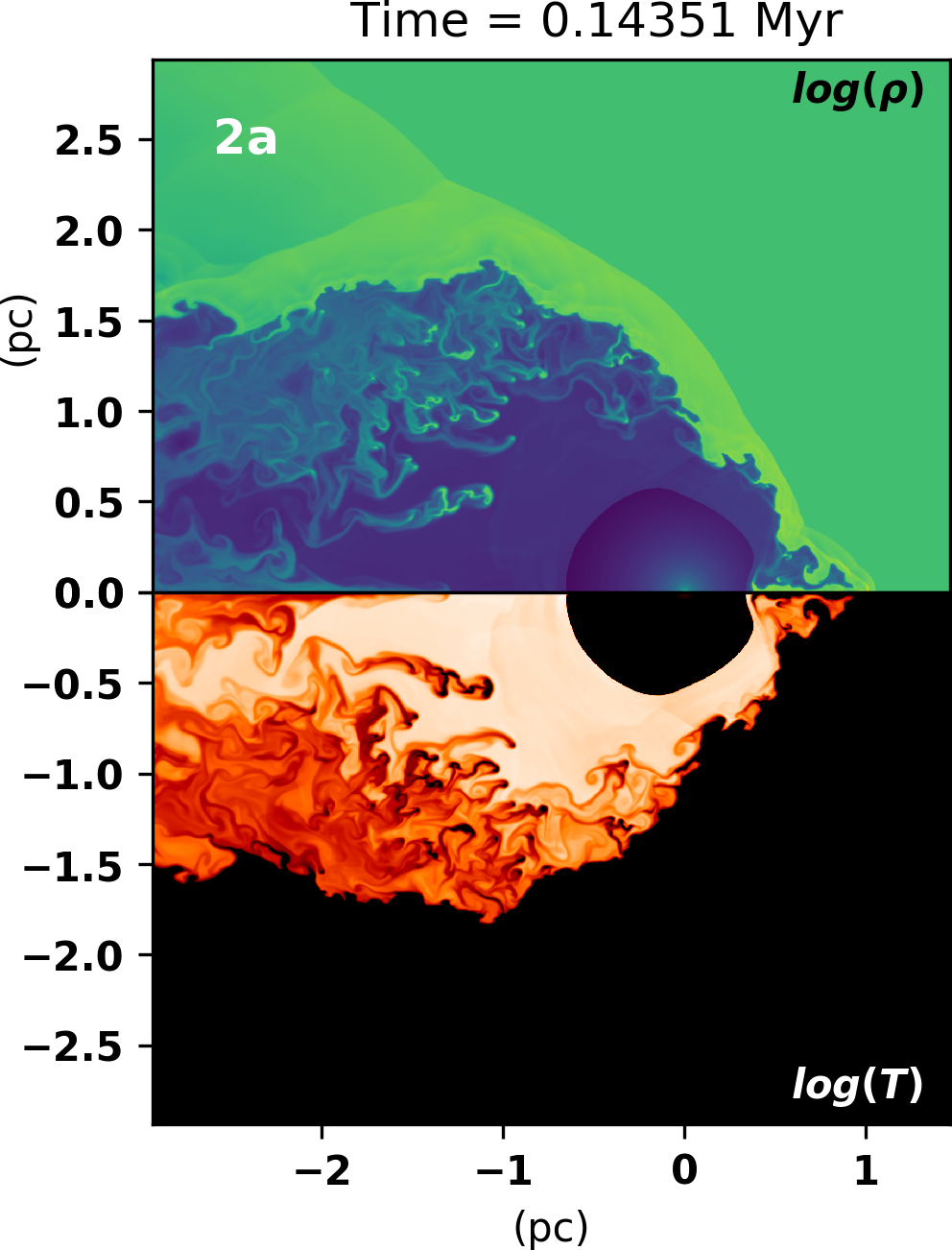}
	\includegraphics[height=.410\textwidth]{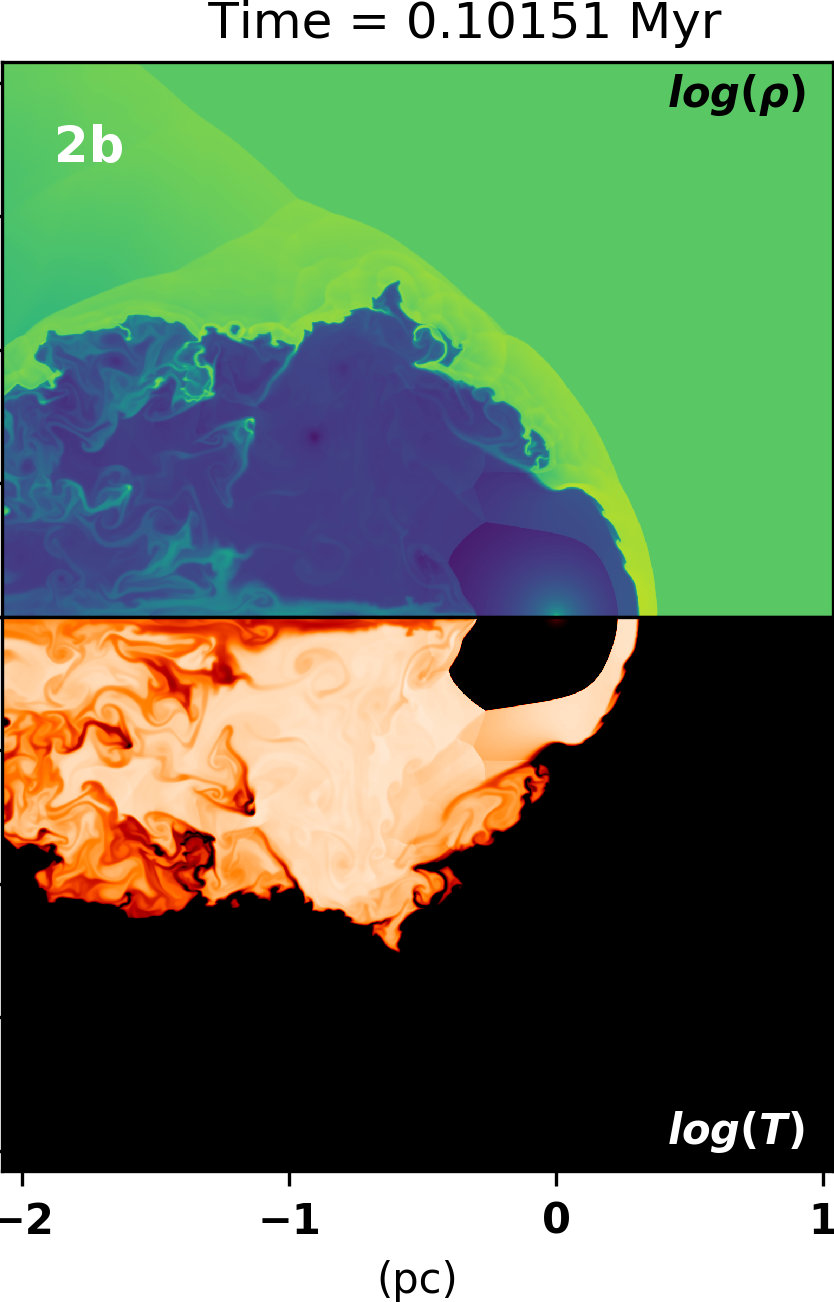}
	\includegraphics[height=.410\textwidth]{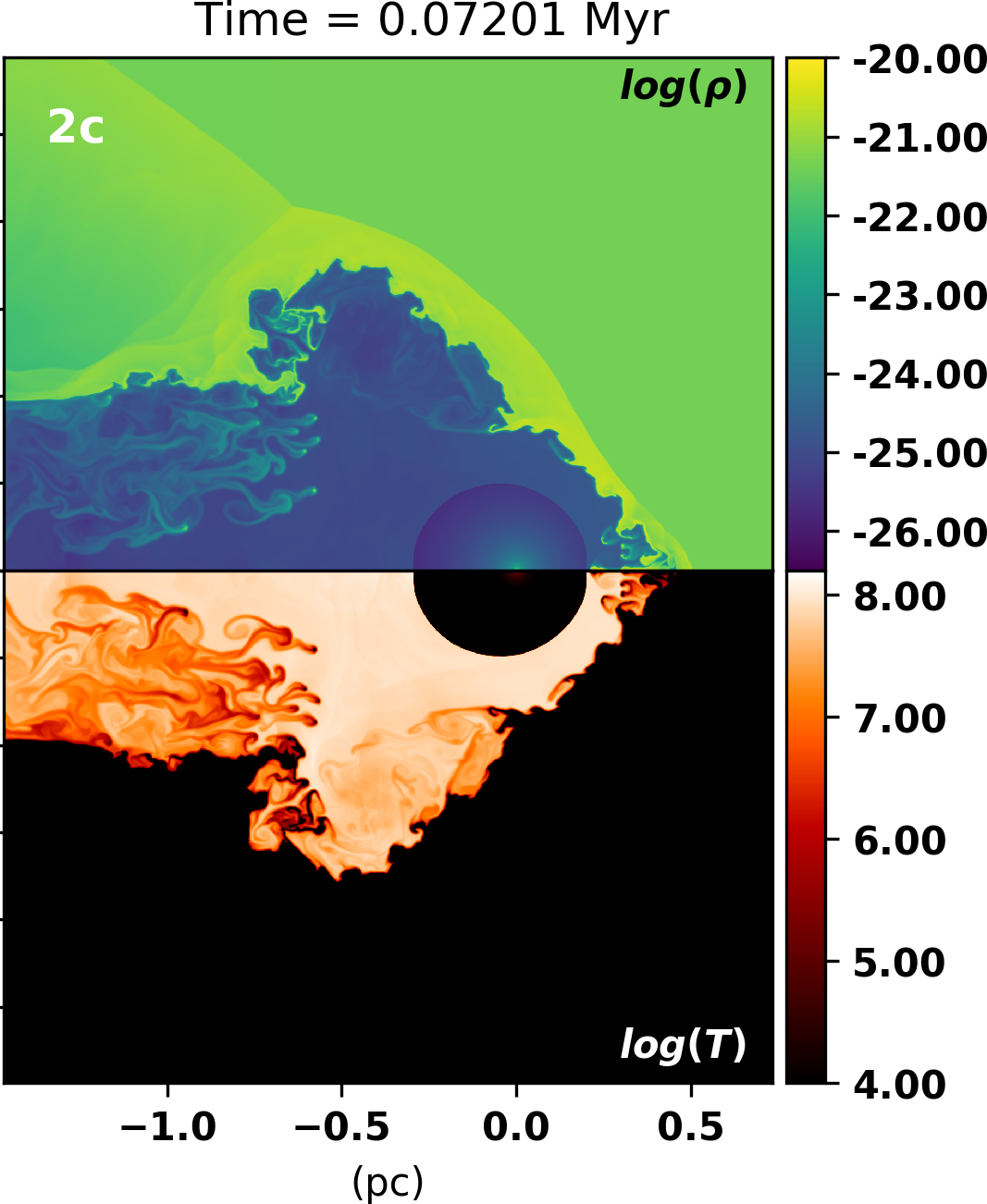} \\
	\includegraphics[height=.410\textwidth]{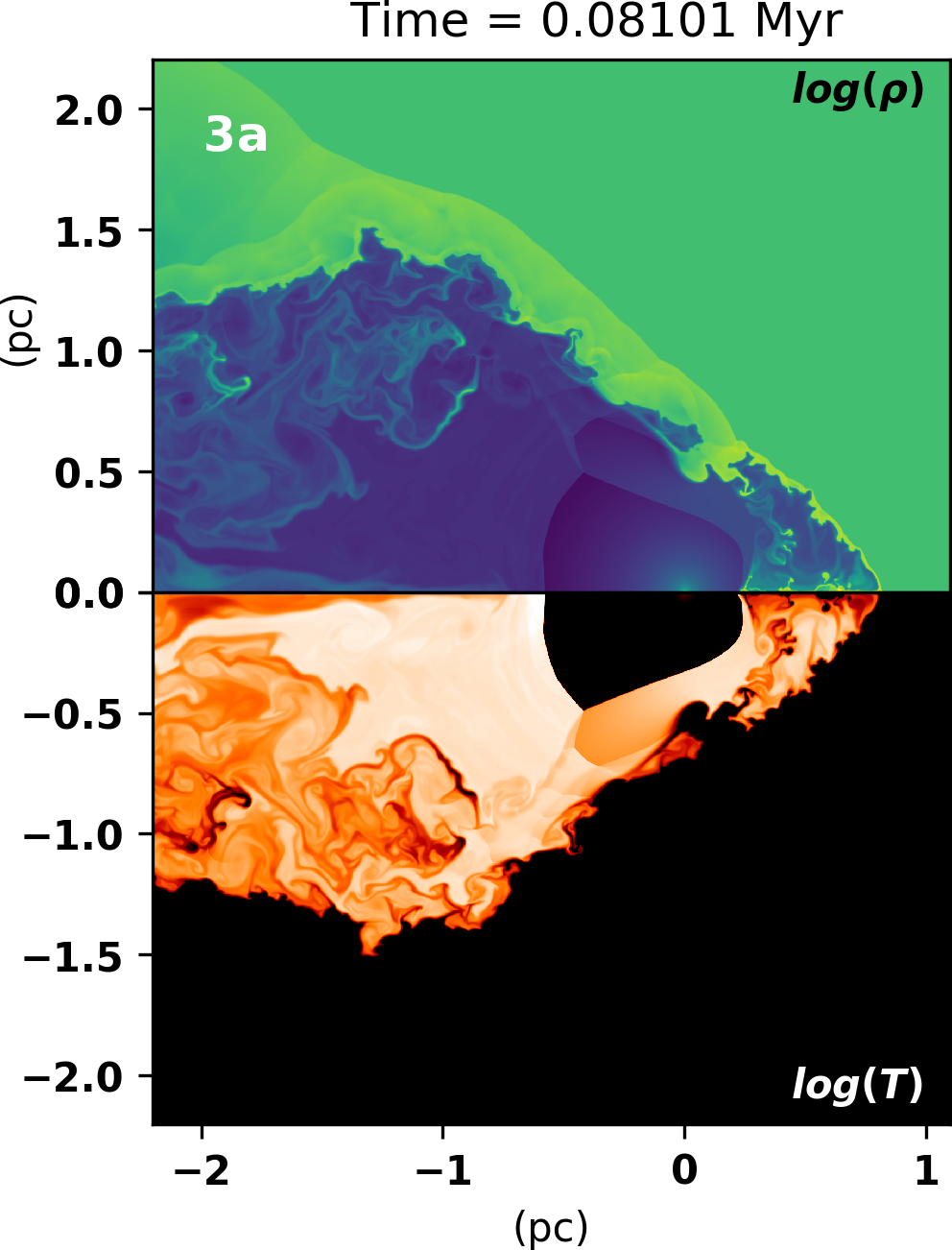}
	\includegraphics[height=.410\textwidth]{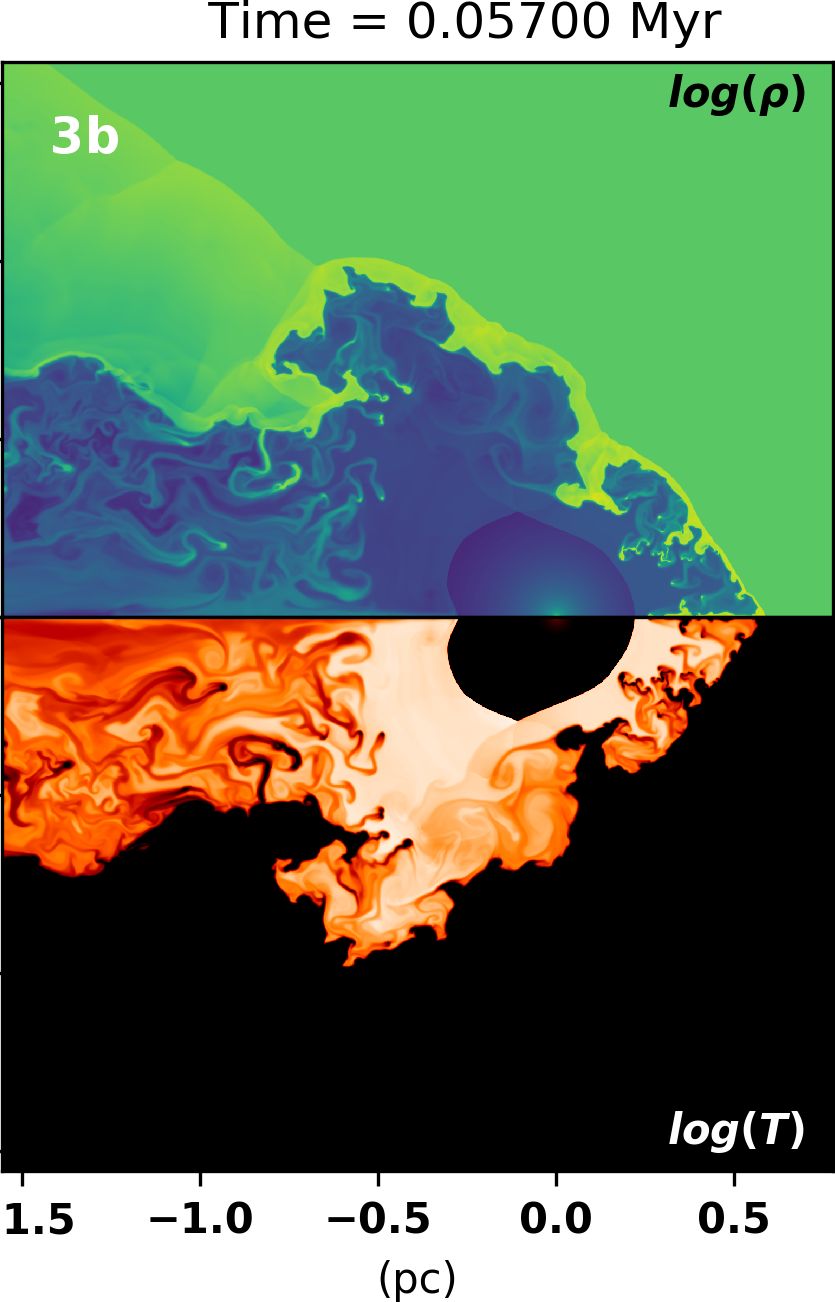}
	\includegraphics[height=.410\textwidth]{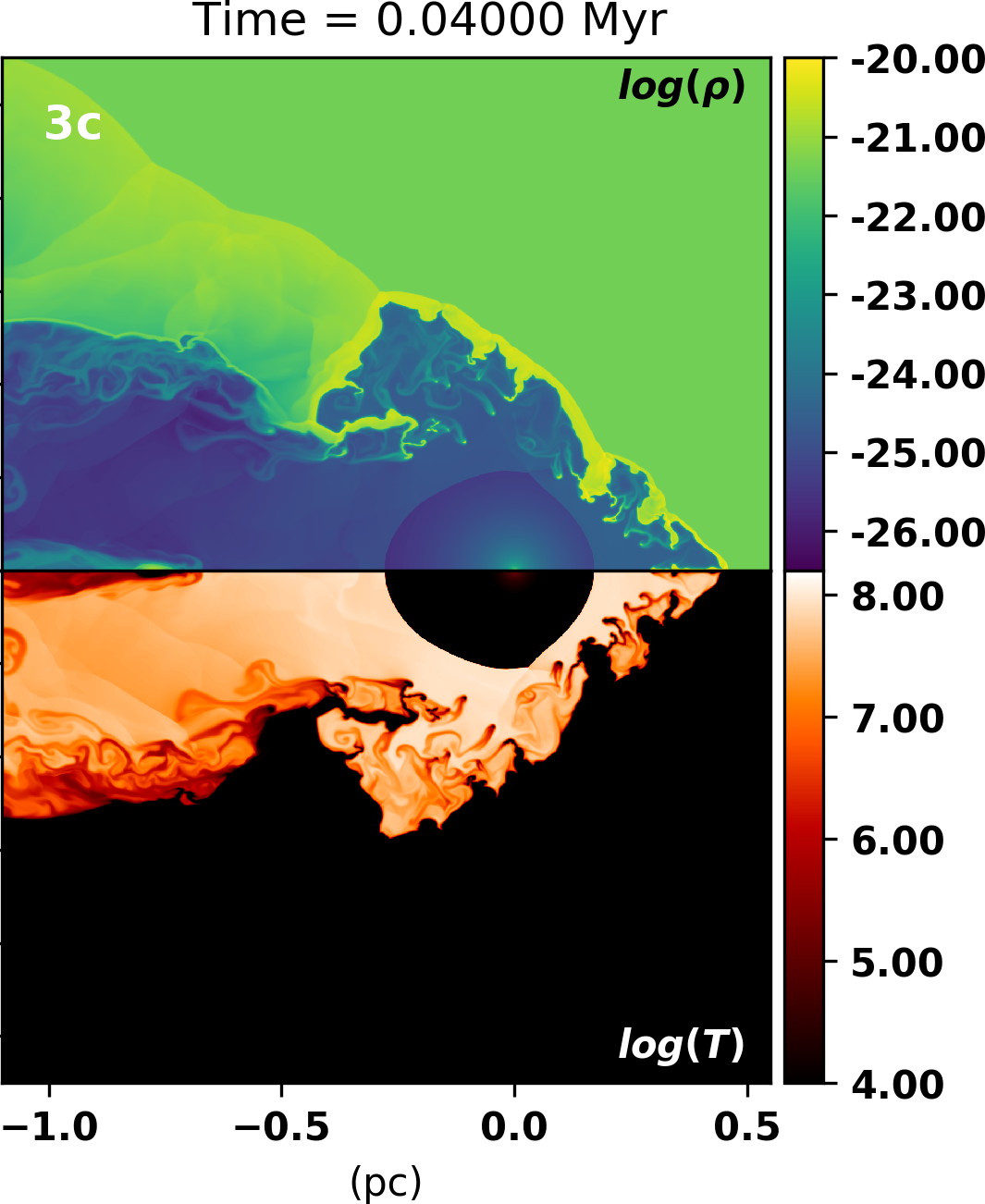}
	\caption{From left to right the panels show a snapshot after 1 crossing time from each simulation in Table \ref{tab:sims}. The top half shows $\log_{10}$ plots of the gas density (g\,cm$^{-3}$) profile of the gas, and the bottom half shows $\log_{10}$ plots of the temperature profile of the gas with white being of higher temperature(K) than black. The star is at the origin. Labels of the different models are shown in white on each panel.}
	\label{sims}	
\end{figure*}

\begin{figure*}[h]
	\centering
	\includegraphics[height=.460\textwidth]{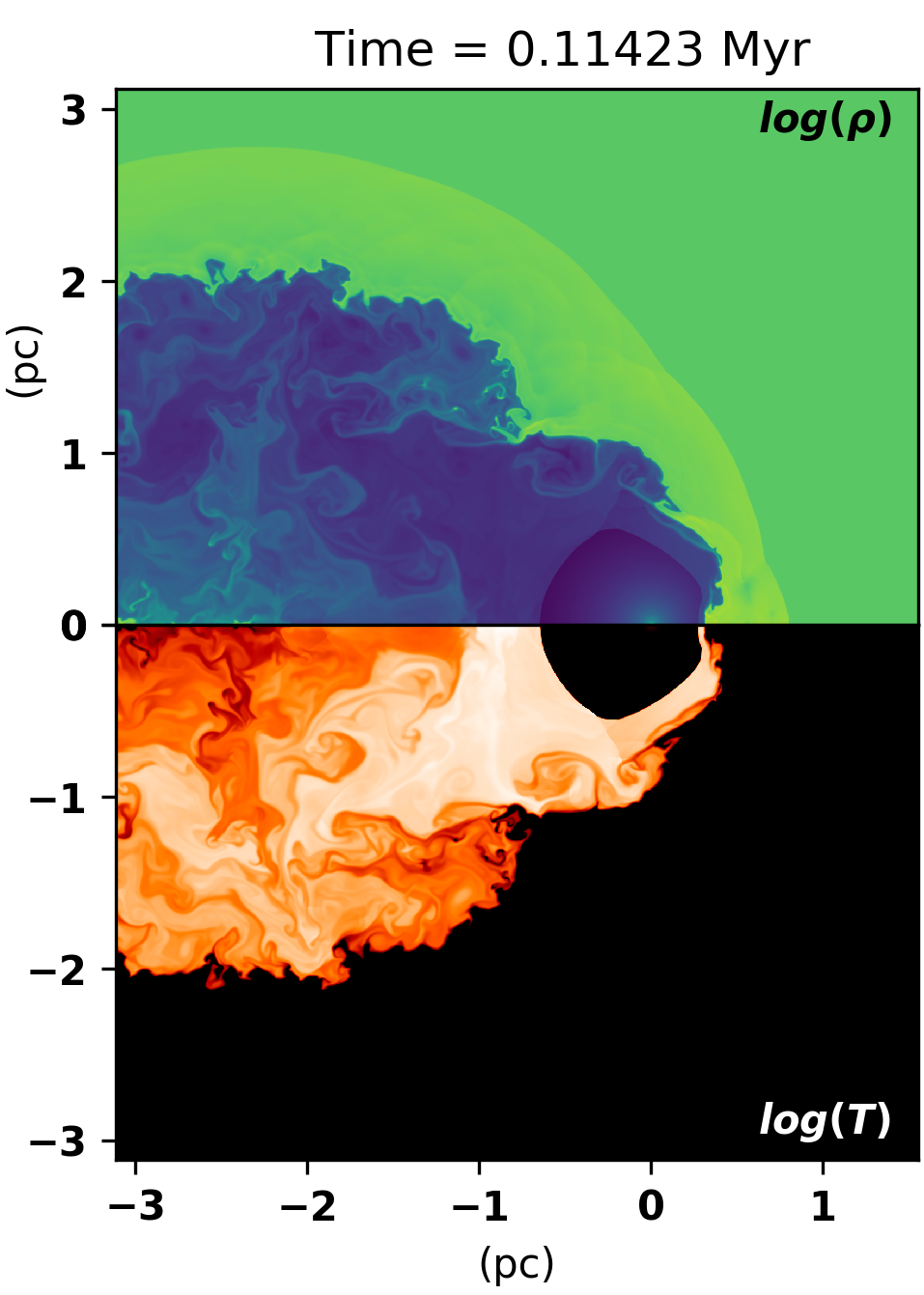}
	\includegraphics[height=.460\textwidth]{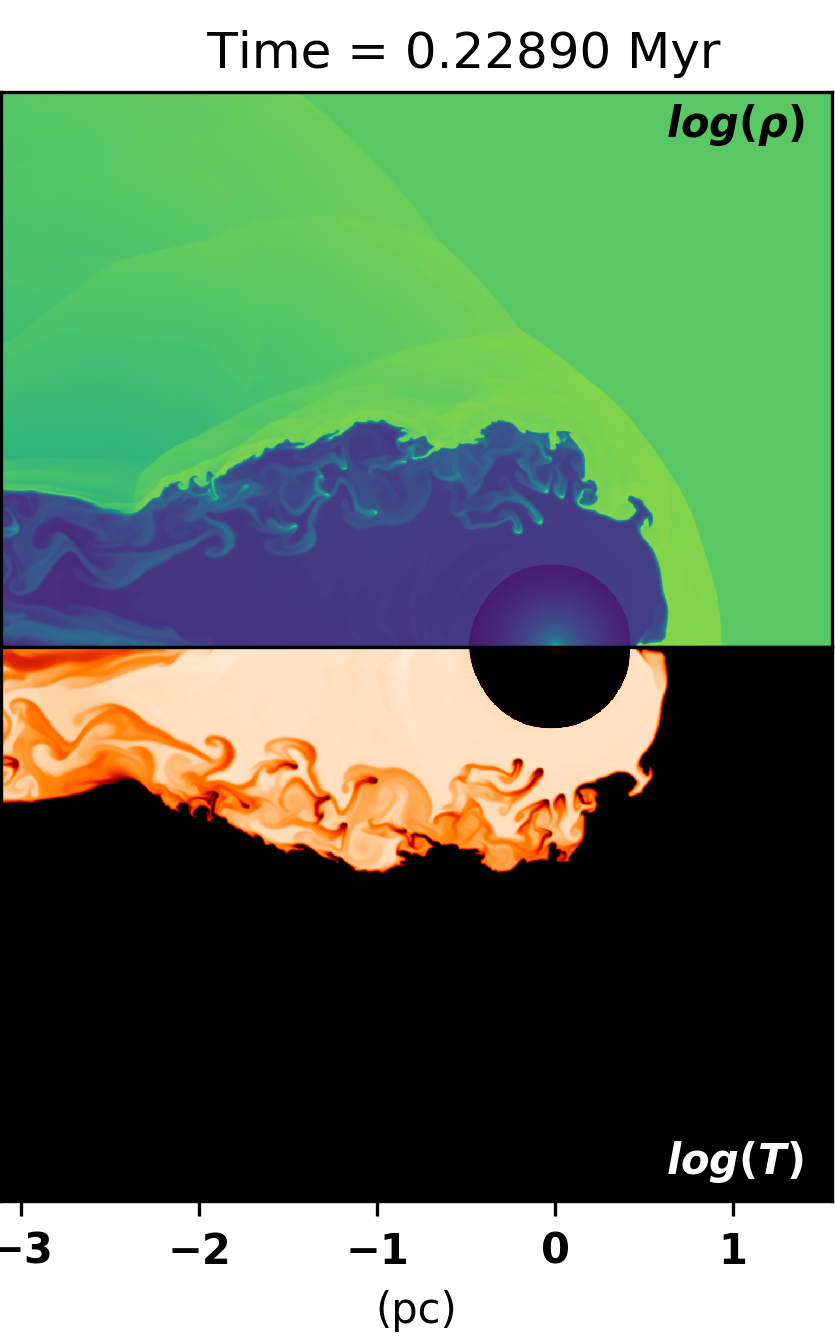}
	\includegraphics[height=.460\textwidth]{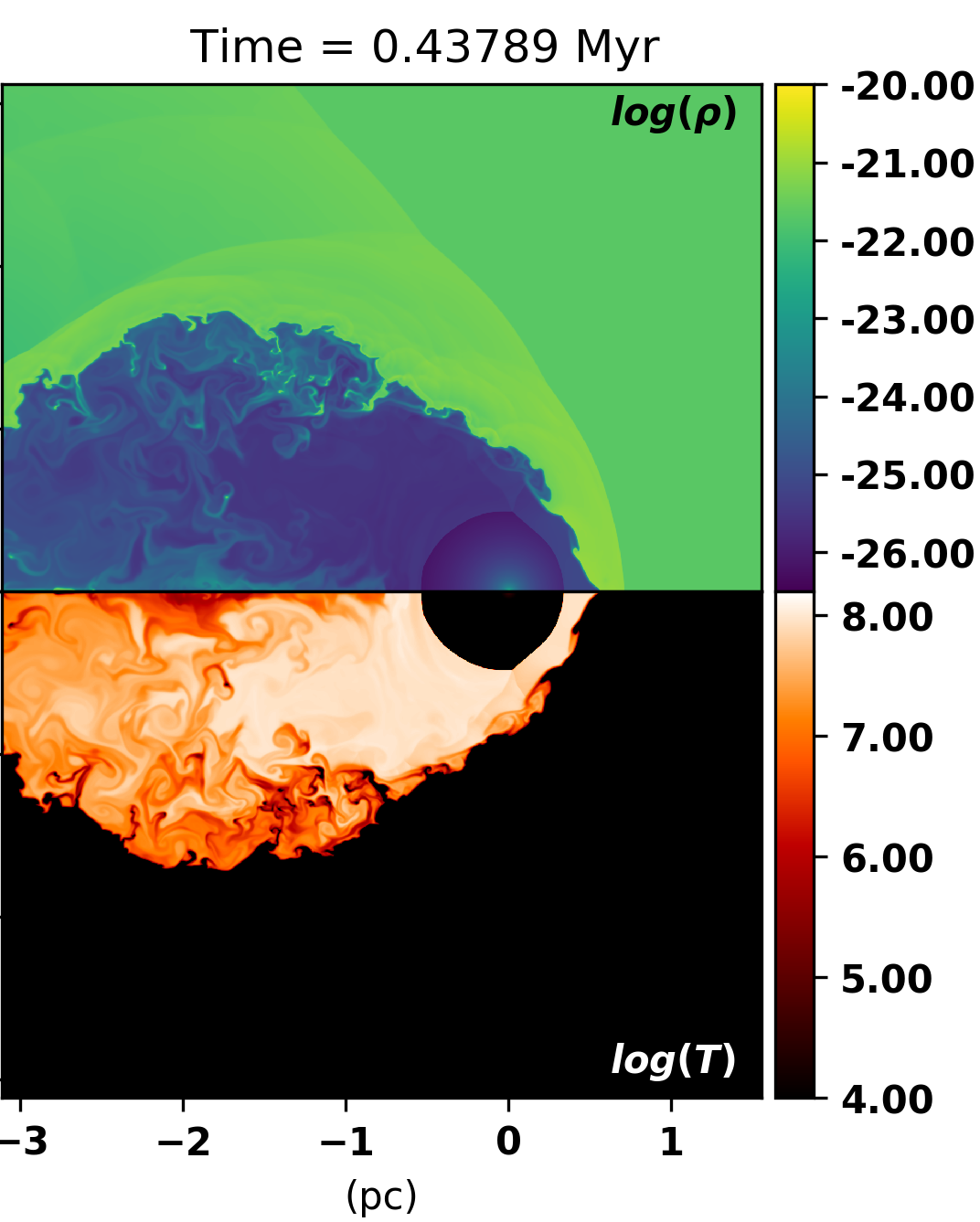}
	
	\caption{Top half shows $\log_{10}$ plots of the gas density (g\,cm$^{-3}$) profile of the gas, and bottom half shows $\log_{10}$ plots of the temperature profile of the gas with white being of higher temperature(K) than black. These plots are from the 1b simulation data. The star is at the origin. From left to right the panels show the simulation after 0.5, 1, and 2 crossing times, respectively.}
	\label{den}
\end{figure*}

\begin{figure*}
	
	\centering
	\includegraphics[height=.460\textwidth]{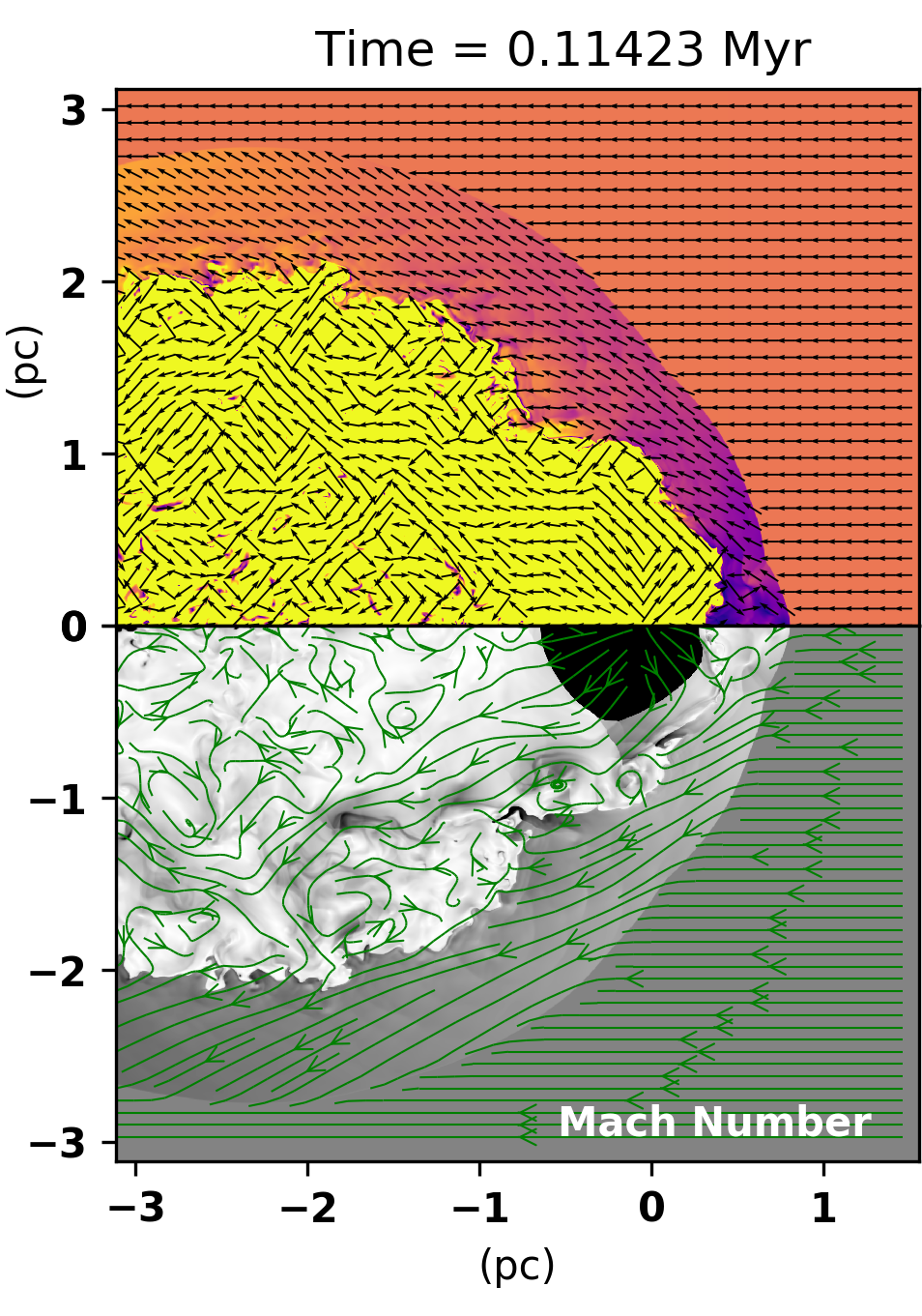}
	\includegraphics[height=.460\textwidth]{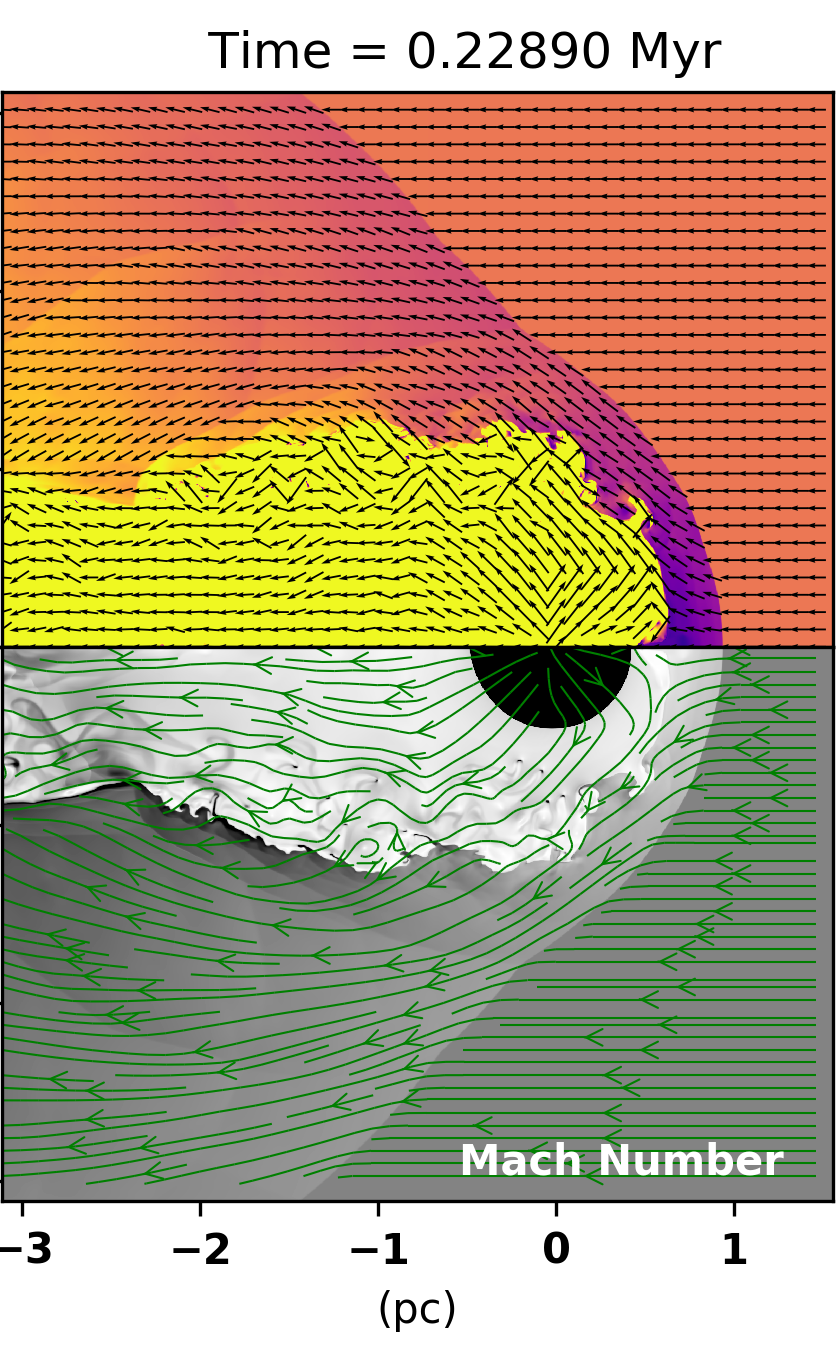}
	\includegraphics[height=.460\textwidth]{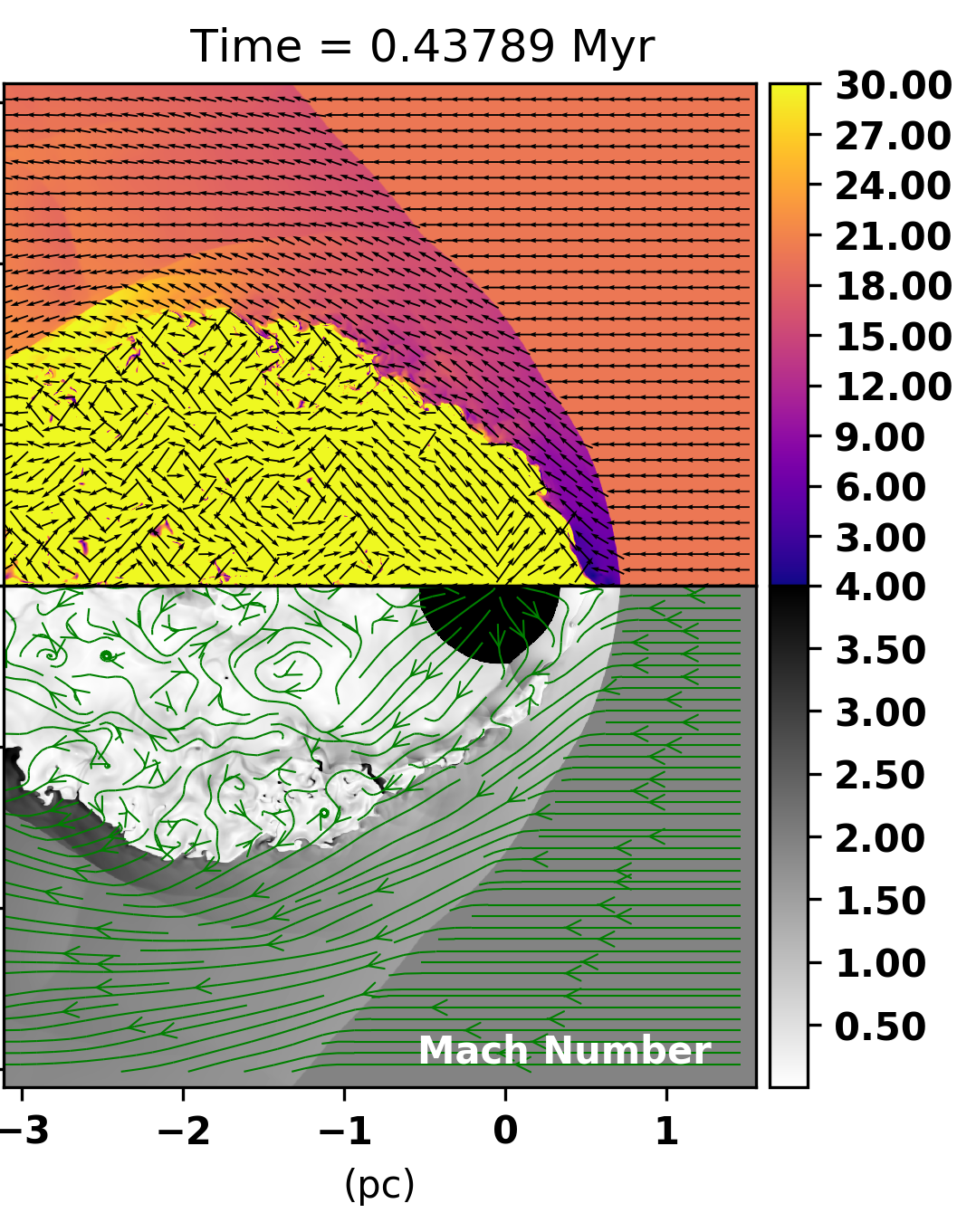}
	
	\caption{Top half plane shows gas velocity (km\,s$^{-1}$) with yellow being of higher velocity than dark purple. Bottom half plane shows the Mach number of the gas with black being of higher Mach number than white. These are from the 1b simulation at the same three time-steps as the density/temperature plots in Fig.~\ref{den}.}
	\label{vel}
	
\end{figure*}

\subsection{Comparison of simulations}

Nine 2D simulations were run using the parameters shown in Table \ref{tab:sims}, all using the stellar wind properties taken from BD +6$0^\circ$\,2522 (see Table \ref{tab:param}). A snapshot after $\sim$ 1 crossing time (time taken for one fluid element to cross the whole simulation box) for each simulation is shown in Fig.~\ref{sims}.
Simulation 2a, 2c, 3a, 3b, and 3c could not be used due to the development of an unstable bow shock and gas pile-up at the apex, which is a well known limitation of 2D simulations \citep[e.g.][]{2014meyer}. The apex is where the stellar wind and the ISM collide head-on, resulting in a stagnation point of the flow. We compared the remaining four simulations to the Bubble Nebula and found that for simulation 1a the bubble was too big and too faint. Simulations 1c and 2b produced nebulae that were too bright. The 1b model was chosen to be the best simulation to compare with the Bubble Nebula because it evolved into an elliptical shape with a smooth bow shock and its synthetic emission maps had similarity to the observational data.

\subsection{Simulation 1b} 
Snapshots from simulation 1b are shown in Fig.~\ref{den}. The snapshots are after 0.5 (0.11\,Myr), 1 (0.23\,Myr), and 2 (0.44\,Myr) crossing times. The upper half plane in the plots shows $\log_{10}$ of the gas density profile. The colour-bar on the right of each plot shows that yellow is the highest density ($10^{-20}$ g\,cm$^{-3}$) and dark blue/purple is the lowest density ($10^{-26}$ g\,cm$^{-3}$). The lower half plane in the plots shows $\log_{10}$ of the gas temperature profile. In this case, white is the highest temperature ($10^8$\,K) and black is the lowest temperature ($10^4$ K).

Fig.~\ref{vel} shows snapshots taken from the 1b simulation at the same crossing times as Fig.~\ref{den} but plotting gas velocity and the Mach number of the gas in a reference frame where the star is at rest. The upper half plane is a plot of the gas velocity profile of the gas in $\kms$. The maximum velocity shown is $30 \, \kms$ so that the velocity gradients in the bow shock are visible. The velocity of the ISM is initially set to $-20 \, \kms$. The wind bubble has much higher velocity gas, comparable to the wind speed of $2500 \, \kms$. Overlayed are vector arrows showing the direction of the flows. These are used to show the position on the bow shock and how the gas from the ISM is swept backwards around the bubble. The lower half plane is a plot of the isothermal Mach number, $\mathcal{M}$ of the gas flow with respect to the star, defined by $\mathcal{M} = |\boldsymbol{v}|/\sqrt{p/\rho}$. Overlayed are streamlines showing the direction of the gas.

Early in its evolution, the bow shock is expanding and is not in equilibrium. Later on, the bow shock begins to approach equilibrium where the total pressure (ram pressure + thermal pressure) is constant. However, there is an exception to this where instabilities disturb the flow. We will now discuss the simulation results where the bow shock is in its equilibrium state. 

At the apex of the bow shock, Fig.~\ref{den} shows that the density is $\sim 10^{-21}$ g\,cm$^{-3}$. The density stays roughly at this value throughout the entire simulation. The compression factor is largest at the apex because the Mach number (1.94) of the shock is biggest there (dismissing the Mach number of the wind directly from the star). The temperature at the bow shock's apex is also in equilibrium with the rest of the ISM ($\sim 10^{4}$\,K). The forward shock in the ISM is basically isothermal because the density is high, the post-shock temperature is $\sim 10^4$\,K, cooling is strong, and the cooling time is very short. Fig.~\ref{vel} shows that at the apex of the bow shock the velocity of the gas is $0 \, \kms$. 

Inside the hot bubble, the gas density is as low as $\sim 10^{-26}$ g\,cm$^{-3}$ and as high as $\sim 10^{-24}$ g\,cm$^{-3}$. The reverse shock in the wind is adiabatic because the density is low, the post-shock temperature is $\sim 10^8$\,K, cooling is weak, and the cooling time is very long. The velocity of the gas reaches $30 \, \kms$ (and certainly higher because the unshocked part of the stellar wind is moving at $2500 \, \kms$). 


The part of the bow shock at 45\degr \, to 90\degr \, from the apex (measured from the star) is where the bow shock expands and the density decreases slowly as the velocity increases.
Densities and temperatures of this region are similar to the apex of the bow shock with high densities (compared to the inside of the bubble) and temperatures are the same with the surrounding ISM. The velocity of the gas here increases as you go further from the apex because the velocity vectors of the ISM and wind don't cancel. The further back you go along the bow shock, the closer the velocity gets to the speed of the ISM ($20 \, \kms$, Mach number of 1.94). 

As the bubble evolves, the bow shock relaxes slowly towards its equilibrium shape (parabolic). However, the bubble itself remains in an elliptic shape. In the later snapshots, the density of the bow shock decreases as $R$ increases. The stellar wind bubble in these simulations also has a periodic change in bubble size. The bubble is at its maximum size roughly every 0.18 Myr (and hence minimum size every 0.09 Myr after a maximum). This can be explained with `Vortex Shedding' and will be discussed further in Sect. \ref{sec:xray}. 

\begin{figure}
	\centering
	\includegraphics[width=.46\textwidth]{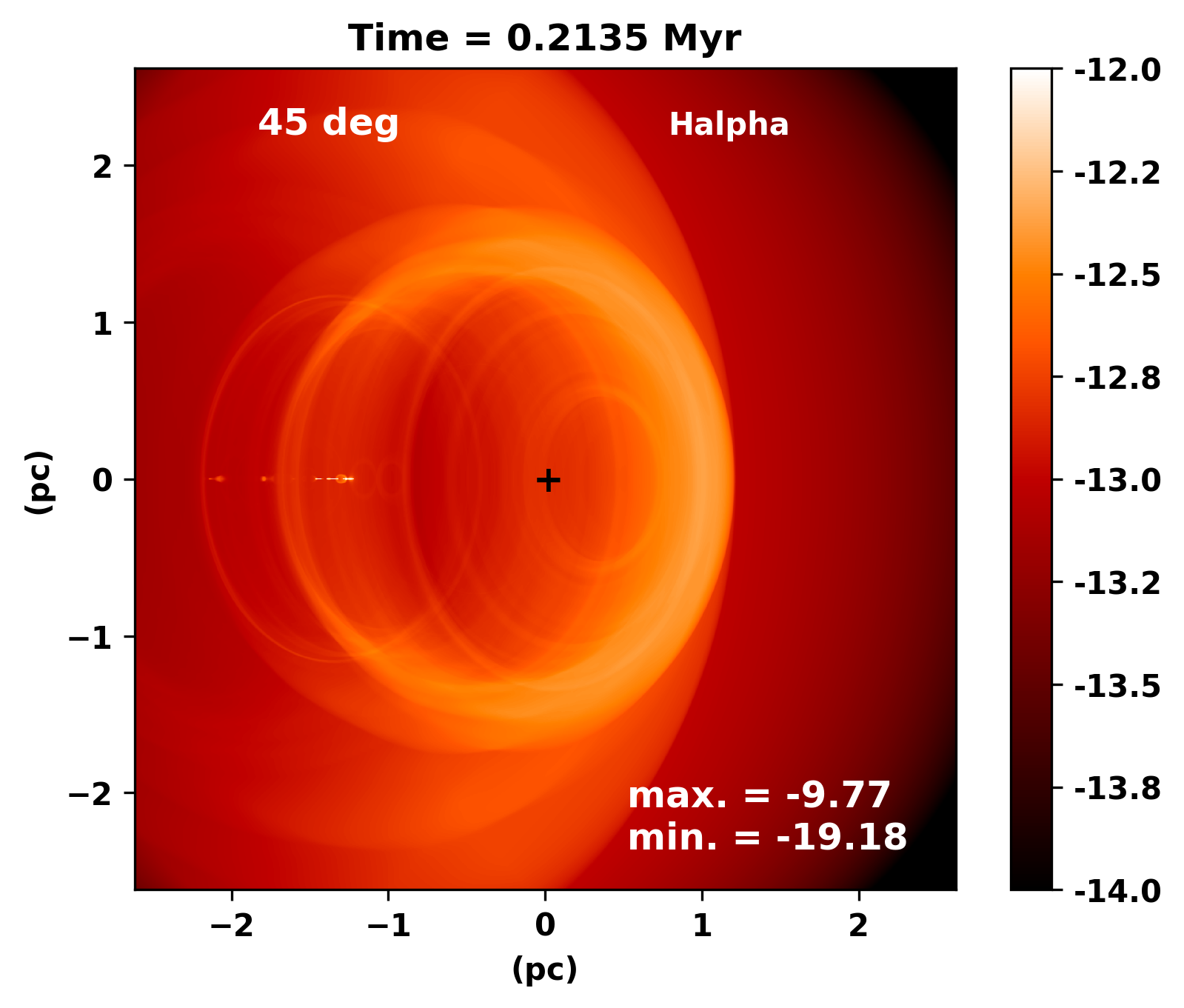} \\
	\includegraphics[width=.46\textwidth]{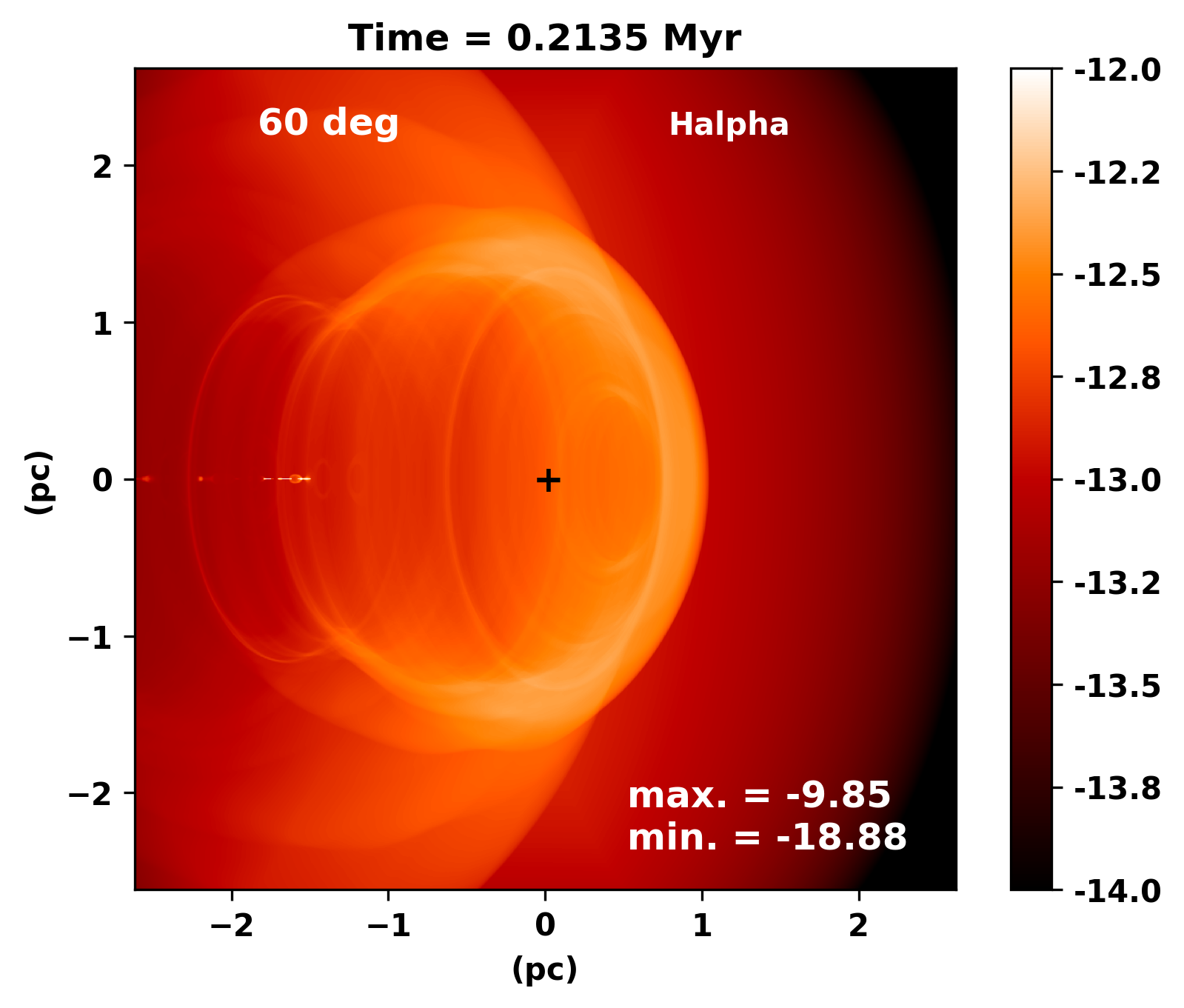} \\
	\includegraphics[width=.47\textwidth]{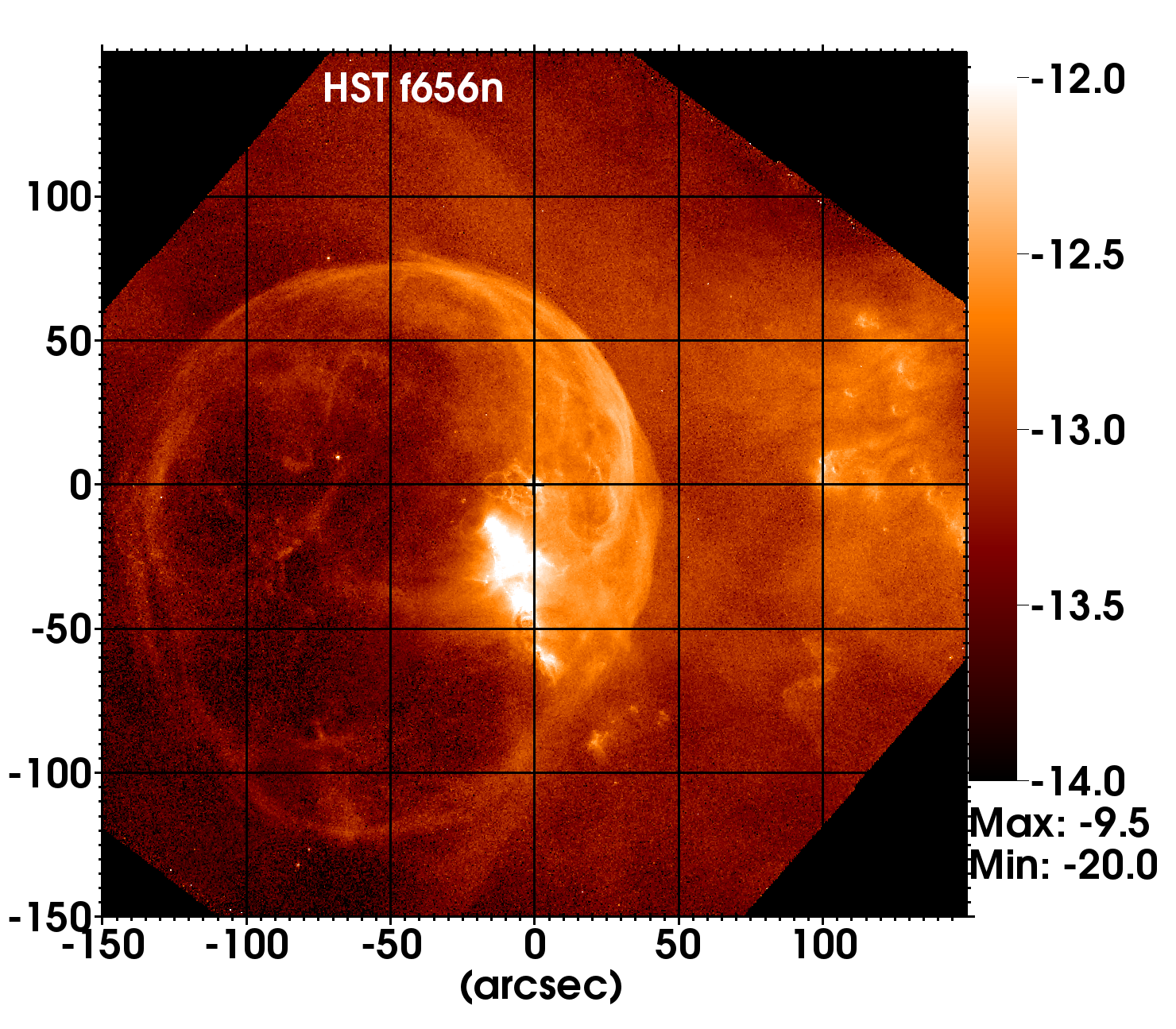}
	
	\caption{Top two images: Synthetic H$\alpha$ emission maps of the bow shock on a logarithmic colour scale (in units of $\mathrm{erg\ cm^{-2}\ s^{-1}\ arcsec^{-2}}$), generated from the 1b simulation for viewing angles of 45\degr \, and 60\degr. Both synthetic images are generated after 0.2135\,Myr of evolution. The coordinates are in pc relative to the star's position. The black cross shows the position of the ionizing star. Bottom plot: \textit{HST} H$\alpha$ (656nm) image of the Bubble Nebula.}
	\label{halpha}
	
\end{figure}

\section{Synthetic emission maps and comparison with observations}
\label{sec:sin}

Synthetic observations are predictions based on theoretical simulations to provide a view of what the astrophysical source in question will appear as to an observer. It is a way of comparing simulations to observational data and to allow the theoretical models to be constrained. \citet{2018NewAR..82....1H} presented a recent detailed description of the power of synthetic observations in star formation and the impact of stars on the ISM. In this section, we present our synthetic optical, IR and X-ray images obtained from our simulations to test the idea that the Bubble Nebula has formed as a result of a bow shock around BD+6$0^\circ$\,2522.

\subsection{Observational data used for comparison}

The H$\alpha$ image that we use for comparison with simulations was obtained by the Hubble Space Telescope (\textit{HST}) (Program Id.: 14471, PI: Zolt Levay) on 2016.02.25. We downloaded the level 2 data products (reduced and calibrated images) from the Mikulski Archive for  Space Telescopes (MAST)\footnote{https://archive.stsci.edu/} at the Space Telescope Science Institute (STScI). The image is composed of 4 tiled WFC3/UVIS pointings with 3 images of 500 seconds exposure per tile (i.e.~1500 seconds exposure per pixel) and was produced by \citet{AviLevChr16} as part of the \textit{Hubble} Heritage Project.
	
The {\it Spitzer} 24\,$\mu$m image of the Bubble Nebula was downloaded from the NASA/IPAC infrared science archive\footnote{http://irsa.ipac.caltech.edu/}. The nebula was observed on 2005.12.05 (Program Id.: 20726, PI: J.\,Hester), using the Multiband Imaging Photometer for {\it Spitzer} (MIPS; \citet{2004ApJS..154...25R}). The image is a level 2 data product, with units of MJy\,sr$^{-1}$ and angular resolution of 6 arcsec.

\begin{figure}
	\centering
	\includegraphics[width=.43\textwidth]{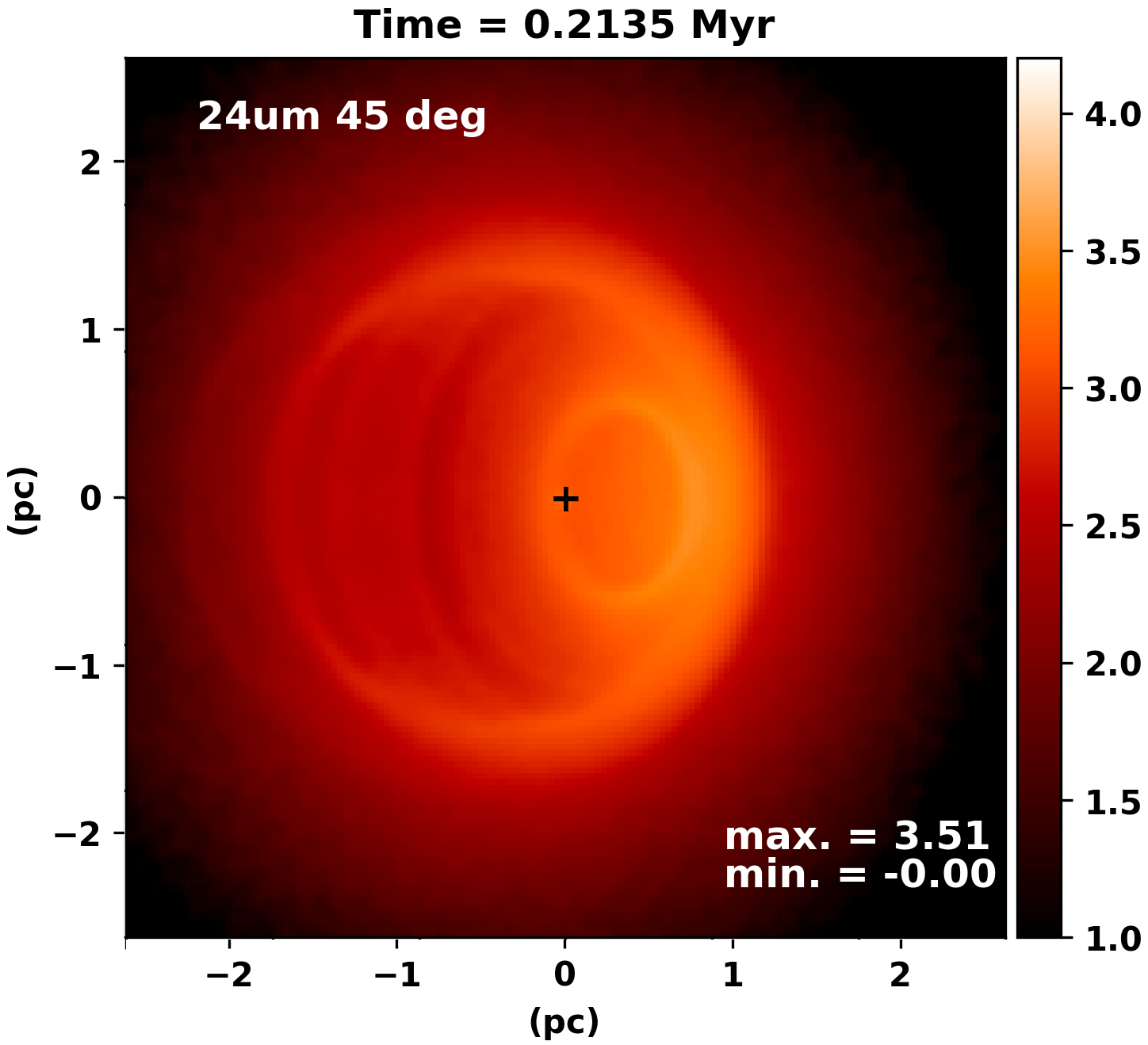} \\
	\includegraphics[width=.43\textwidth]{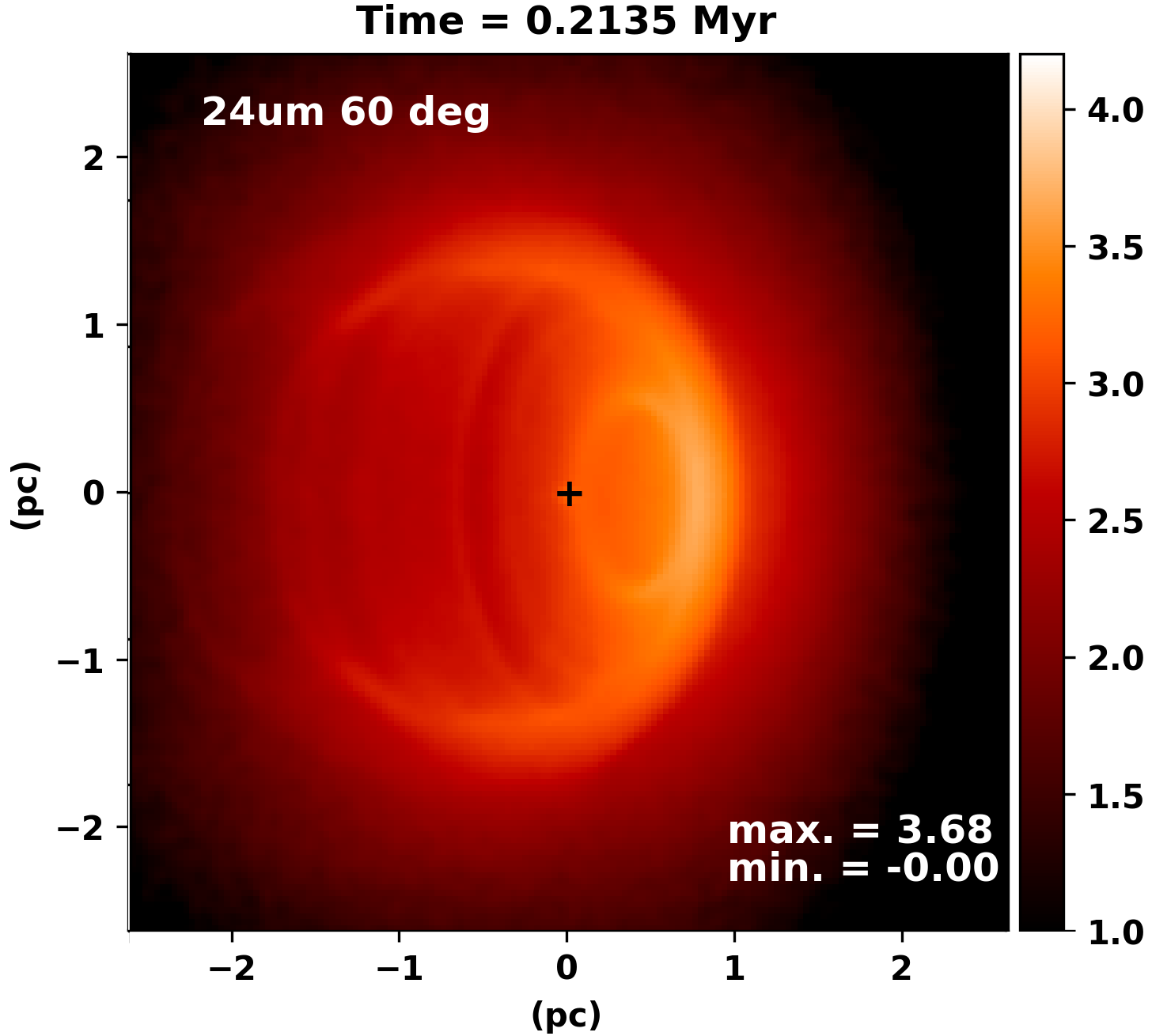} \\
	\includegraphics[width=.47\textwidth]{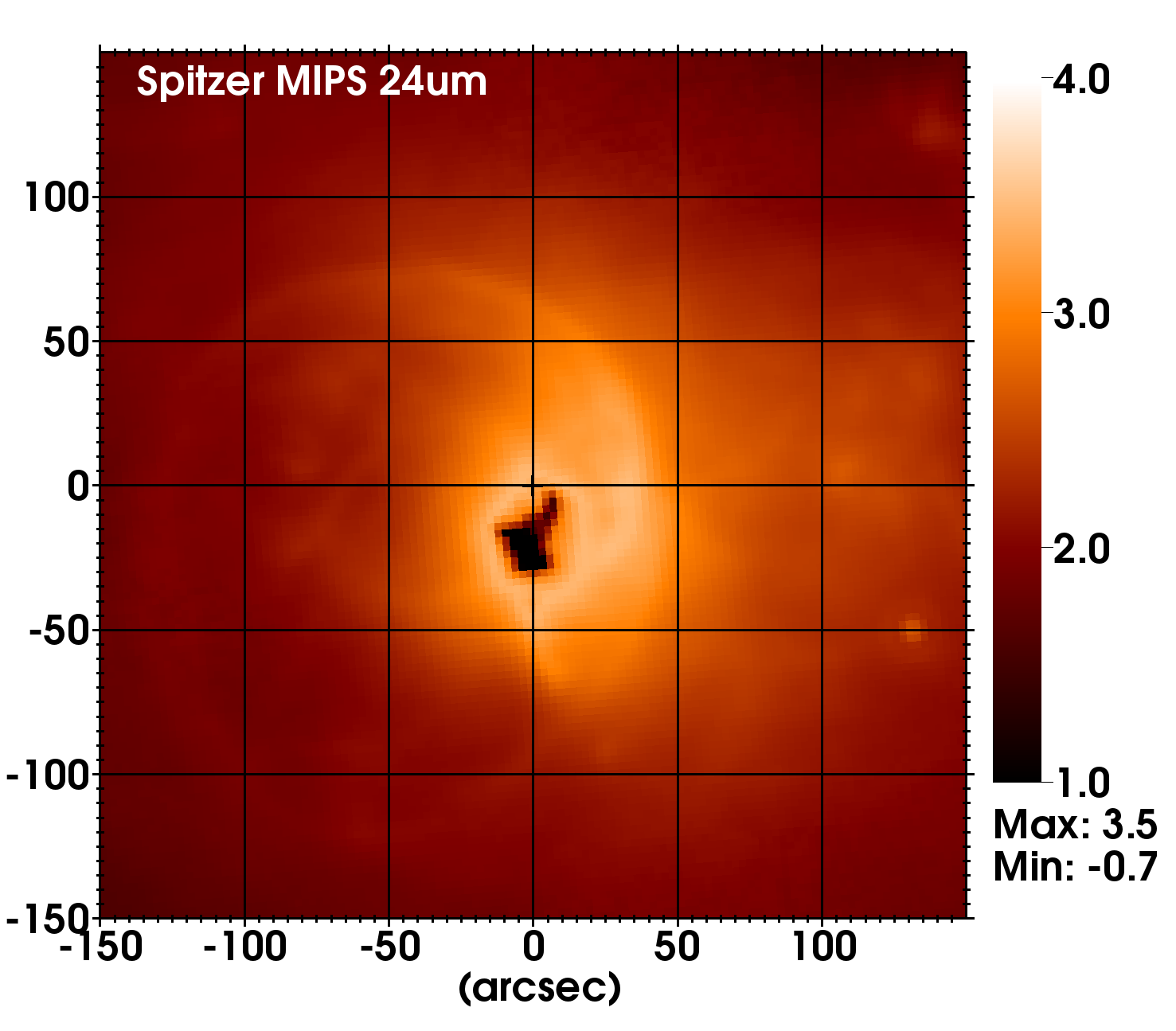}
	
	\caption{Top two images: Synthetic infrared emission maps of the bow shock on a logarithmic colour scale (in units of $\mathrm{MJy\,st er}^{-1}$), generated from the 1b simulation for viewing angle angles of 45\degr \, and 60\degr. Both synthetic images are generated after 0.2135\,Myr of evolution and smoothed to the angular resolution of \textit{Spitzer} of 6 arcsec. The coordinates are in pc relative to the star's position. The black cross shows the position of the ionizing star. Bottom plot: \textit{Spitzer} 24\,$\mu$m image of the Bubble Nebula.}
	\label{spitzer}
	
\end{figure}

\subsection{Calculating H$\alpha$ emission}

Our 2D cylindrical models (in the $R-z$ plane) are rotationally symmetric about the $z$-axis. We utilize this symmetry to produce synthetic emission maps of the 3D structure. We developed a raytracing method to calculate synthetic images described in Appendix\,B, using the symmetry of the simulation to generate projection through 3D space at an angle to the grid (similar to \citet{2006ApJS..165..283A} for radio emission). Synthetic H$\alpha$ emission maps were generated using this method. The H$\alpha$ emissivity was calculated by interpolating a table in \citet{Ost89} as 

\begin{equation}
j_\mathrm{H\alpha} = 2.63\times10^{-33} \frac{n_\mathrm{e} n_\mathrm{H}}{T^{0.9}}\,\mathrm{erg\,cm^{-3}\,s^{-1}\,arcsec^{-2}} \,.
\nonumber
\end{equation}

Fig.~\ref{halpha} compares synthetic H$\alpha$ emission maps taken from the 1b simulation with the \textit{HST} H$\alpha$ image. We rotated the \textit{HST} image so that the $x$-axis is along the direction of stellar motion as suggested by the \textit{Gaia} DR2 data \citep{2018A&A...616A...1G}. These images show the H$\alpha$ brightness of the Bubble Nebula with units of ${\rm erg} \, {\rm cm}^{-2} \, {\rm s}^{-1} \, {\rm arcsec}^{-2}$ on a logarithmic scale.

The H$\alpha$ emission therefore traces the densest parts of the bow shock and the forward shock is clearly visible in Fig.~\ref{halpha}. At the apex of the bow shock there is a high H$\alpha$ brightness with intensity $10^{-12}$ erg\,cm$^{-2}$\,s$^{-1}$\,arcsec$^{-2}$. Even though Fig.~\ref{den} shows that the bubble is not closed in the negative $z$-direction, in H$\alpha$ it looks like a closed bubble because of projection effects. The brightest pixels in the \textit{HST} image are from a bright-rimmed cloud near the wind bubble but the brightness of the bubble itself is comparable to the synthetic images. 
 
\begin{figure*}[htp]
	
	\includegraphics[width=.138\textwidth]{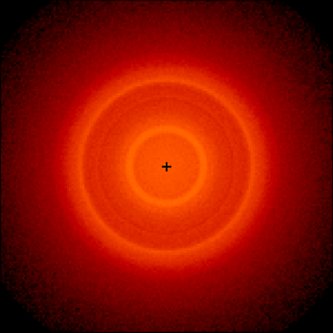}
	\includegraphics[width=.138\textwidth]{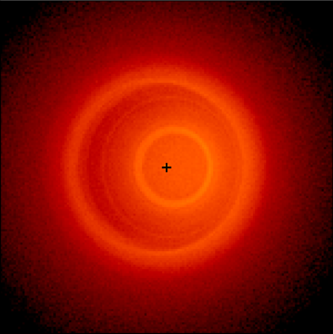}
	\includegraphics[width=.138\textwidth]{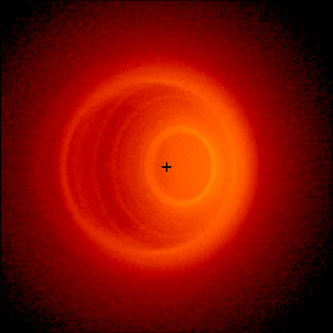}
	\includegraphics[width=.138\textwidth]{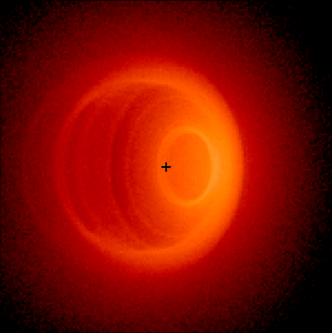}
	\includegraphics[width=.138\textwidth]{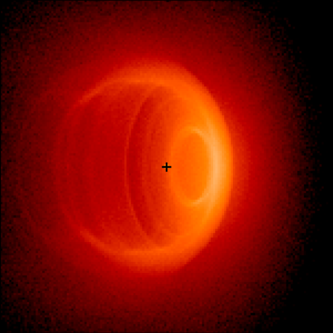}
	\includegraphics[width=.138\textwidth]{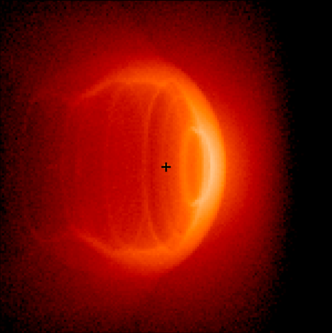}
	\includegraphics[width=.138\textwidth]{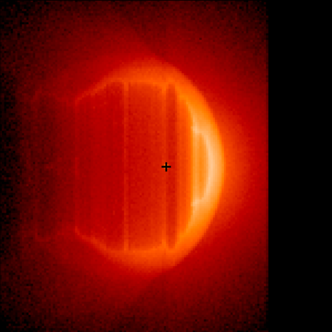} \\
	\includegraphics[width=.138\textwidth]{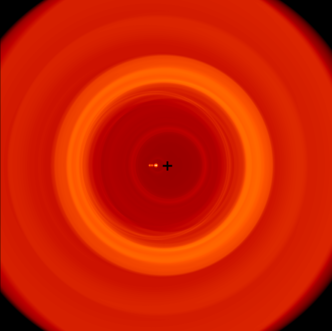}
	\includegraphics[width=.138\textwidth]{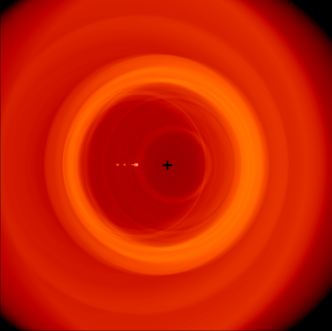}
	\includegraphics[width=.138\textwidth]{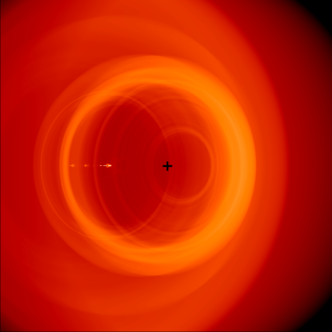}
	\includegraphics[width=.138\textwidth]{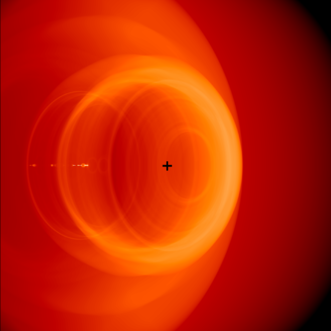}
	\includegraphics[width=.138\textwidth]{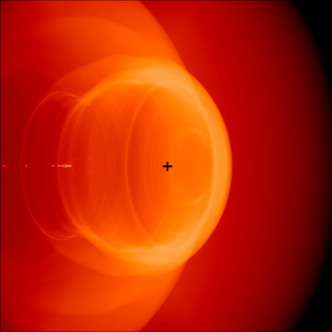}
	\includegraphics[width=.138\textwidth]{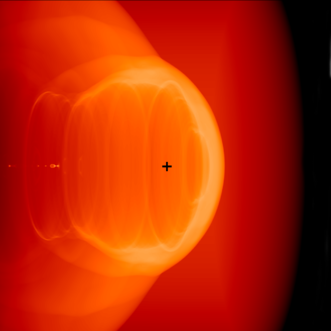}
	\includegraphics[width=.138\textwidth]{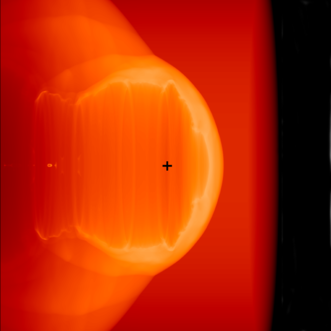} \\
	\includegraphics[width=.138\textwidth]{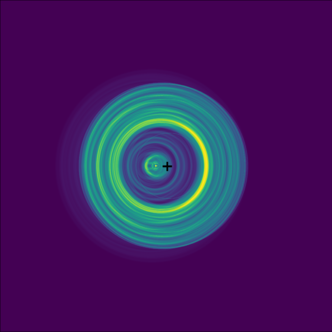}
	\includegraphics[width=.138\textwidth]{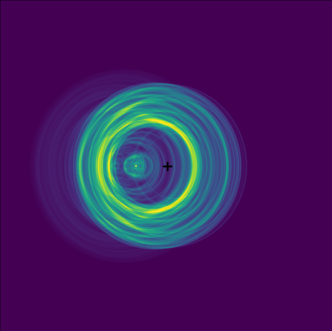}
	\includegraphics[width=.138\textwidth]{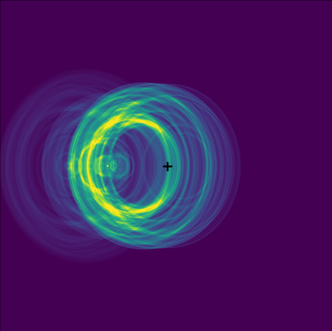}
	\includegraphics[width=.138\textwidth]{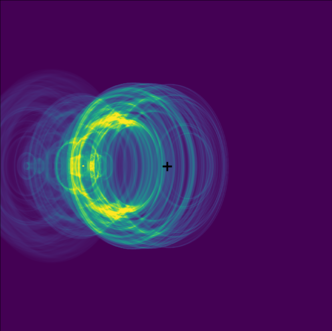}
	\includegraphics[width=.138\textwidth]{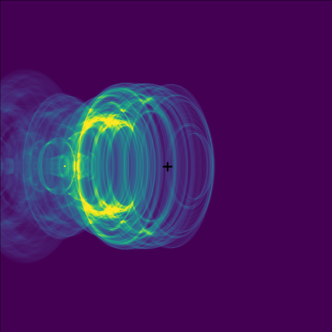}
	\includegraphics[width=.138\textwidth]{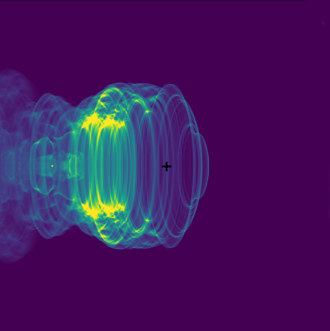}
	\includegraphics[width=.138\textwidth]{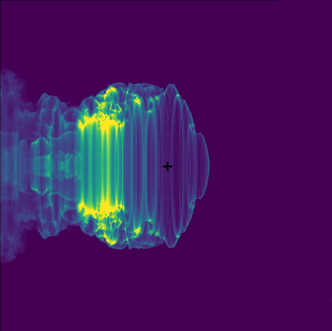} \\
	\includegraphics[width=.138\textwidth]{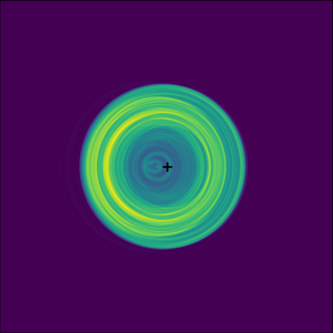}
	\includegraphics[width=.138\textwidth]{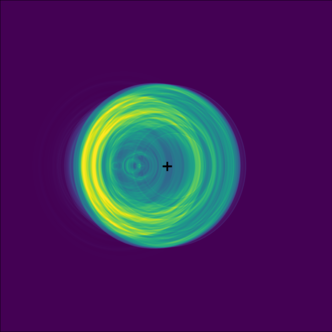}
	\includegraphics[width=.138\textwidth]{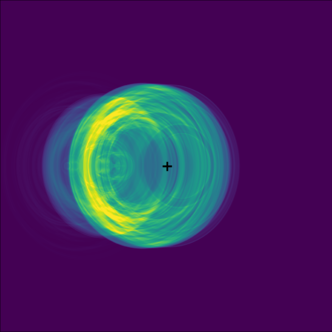}
	\includegraphics[width=.138\textwidth]{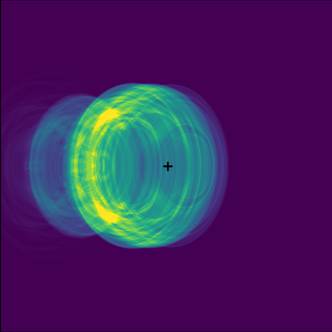}
	\includegraphics[width=.138\textwidth]{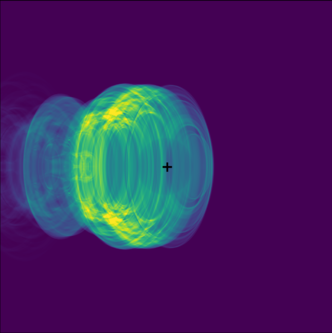}
	\includegraphics[width=.138\textwidth]{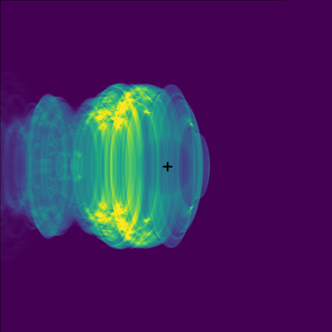}
	\includegraphics[width=.138\textwidth]{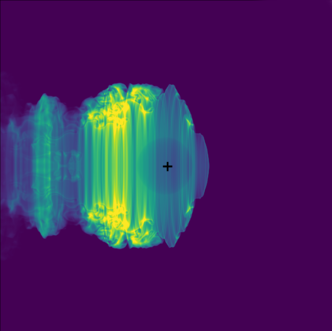}
	
	\caption{A snapshot (after 1 crossing time) from the 1b simulation was used to generate synthetic images of the Bubble Nebula at angles from $0^{\circ} - 90^{\circ}$ with respect to the direction of stellar motion. The first row shows the 24$\mu$m emission, second row shows the H$\alpha$ emission, third row shows the soft X-ray emission, and the fourth row shows the hard X-ray emission. From left to right, the angles are $0^{\circ}$, $15^{\circ}$, $30^{\circ}$, $45^{\circ}$, $60^{\circ}$, $75^{\circ}$, and $90^{\circ}$. For the H$\alpha$ and X-ray image we plot $1^{\circ}$ and $89^{\circ}$ instead of $0^{\circ}$ and $90^{\circ}$ because of technical limitations. Each image is centred on the star (black cross) and is  roughly 5$\times$5 parsec.}
	\label{angles}
\end{figure*}

We made synthetic images at angles from 0\degr \, to 90\degr \, (with 15\degr \, step) between the line of sight and the velocity vector of the star. Visual inspection showed that 45\degr \, and 60\degr \, were most similar to observations. The other angles can be seen in Fig.~\ref{angles}, where we show synthetic emission maps in 24$\mu$m, H$\alpha$, soft and hard X-rays at angles of $0^{\circ} - 90^{\circ}$. The criteria we used were the downstream brightness compared with the apex, and the positioning of the wings of the bow shock on the top and bottom of the nebula. Supporting our visual estimate, the position of the star in the nebula follows the 60\degr \, synthetic snapshot (see Fig.~\ref{spitzer_position}). Both images also show a high H$\alpha$ brightness at the apex of the bow shock, the predicted brightness is quantitatively consistent, and the physical appearance of the nebula at the apex is similar. The rings are an artifact of the symmetry of the 2D simulation.

\subsection{Calculating dust/infrared emission with \textsc{TORUS}}
\label{sec:torus}

We simulate dust emission maps from our models by post processing them using the Monte Carlo radiation transport and hydrodynamics code \textsc{TORUS} (e.g. \citet{2000MNRAS.315..722H}, \citet{2004MNRAS.351.1134K}, \citet{2015MNRAS.448.3156H}). The procedure here is very similar to that described in \citet{2016A&A...586A.114M}. Snapshots from the \textsc{pion} calculations are saved in \textsc{fits} format and mapped onto the \textsc{TORUS} grid using a bilinear interpolation for 2D models. Because the \textsc{pion} calculations do not compute the dust temperature we first perform a dust radiative equilibrium temperature calculation before producing synthetic observations. For both the radiative equilibrium and synthetic observation calculations we use a Monte Carlo approach based on photon packet propagation introduced by \citet{1999A&A...344..282L}. Further details of the implementation in \textsc{TORUS} are given in the aforementioned papers. 

To ensure a dust-free wind blown region we remove the dust wherever the temperature is more than 10$^6$\,K. In the remainder of the grid we assume a gas-to-dust mass ratio of 160 \citep{2004ApJS..152..211Z}, which is comprised of 70\% silicates \citep{2003ApJ...598.1026D} and 30\% carbonaceous \citep{1996MNRAS.282.1321Z} grains. We assume that no PAHs survive within the H\,\textsc{ii} region. For both the silicate and carbonaceous grains we assume minimum and maximum grain sizes of 0.005 and 0.25\,$\mu$m respectively. The size distribution itself between these limits is a power law $\textrm{d}n/\textrm{d}a \propto a^\text{-q}$ \citep{1977ApJ...217..425M}, where we take $q=3.3$. For the stellar spectrum we use a \citet{1993PhST...47..110K} spectral model with the same temperature as BD+6$0^\circ$\,2522.

We take the same snapshot from the 1b simulation to be run with the \textsc{TORUS} code. The code produces dust emission maps (24$\mu$m) at angles from 0\degr--90\degr \, for each snapshot. The top two images in Fig.~\ref{spitzer} show the results of the \textsc{TORUS} code for 24$\mu$m emission at angles of 45\degr and 60\degr \, between the line of sight and the velocity vector of the star. The images from the 2D radiative transfer calculations have a length and width of 5 pc and pixel diameter of 0.0195\,pc. 
 
The synthetic snapshots are smoothed to the resolution of \textit{Spitzer} to accurately compare the features in the simulated data with the observational data. This has been achieved by convolving the data and does not account for instrumental response. It can be seen that the maximum brightness of the synthetic snapshots ($10^{3.51}$, $10^{3.62}$ $\mathrm{MJy\,sr}^{-1}$) matches the maximum brightness of the \textit{Spitzer} image ($10^{3.5}$ $\mathrm{MJy\,sr}^{-1}$). The minimum brightness is slightly different because the synthetic maps only simulated the bubble and not the background radiation, whereas \textit{Spitzer} detected background emission in the field of view.

There are distinct morphological similarities between the synthetic images and the observations. The most noticeable similarity would be the spherical emission surrounding the stellar wind bubble which is at its brightest near the apex of the bow shock. The size of the bubble is consistent with the observations in that it has a radius of $\sim$2.5-3\,pc.

\subsection{Star's position in the Bubble Nebula}
\begin{figure}[h]
	\centering
	\includegraphics[width=.46\textwidth]{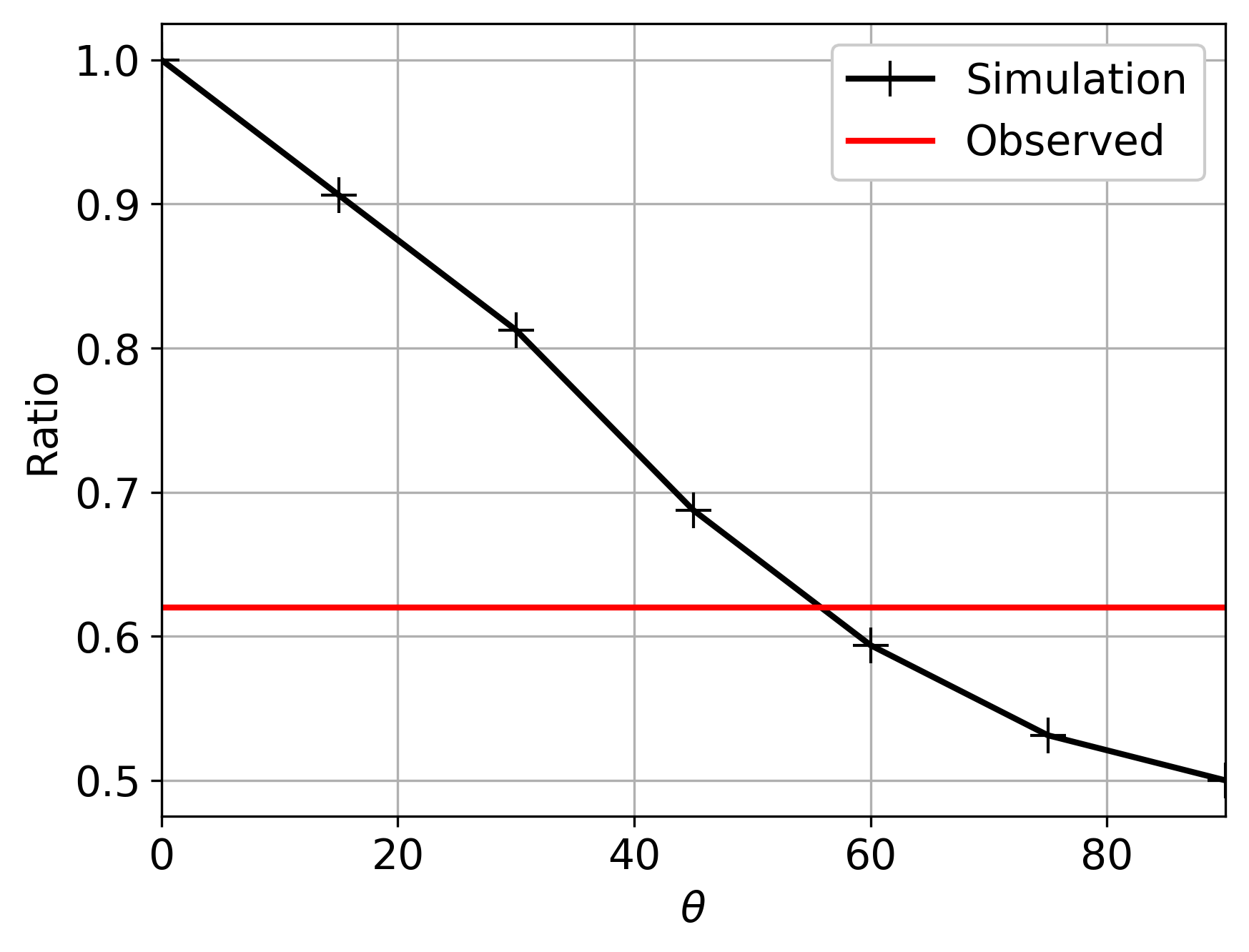}
	
	\caption{Ratio of the stand-off distance $R_\mathrm{SO}$ to the radius of the bow shock at an angle of 90\degr \, from the direction of the stellar motion, as a function of viewing angle, $\theta$. The observed ratio is taken from the \textit{Spitzer} image and simulation results are from the synthetic infrared images.}
	\label{spitzer_position}
	
\end{figure}

Both the H$\alpha$ and infrared synthetic emission maps are quantitatively and qualitatively consistent with the observational data and bow shock interpretation. The position of the central star in the 60\degr \, H$\alpha$ and infrared images match that of the central star in the retrospective observations. This is shown quantitatively with Fig.~\ref{spitzer_position}, where we show the star's position in the nebula as a function of the angle the nebula is rotated about the line of sight. The red line then shows the ratio of the star's position in the observed nebula. Where the red line and the black curve intersects gives an indication of what angle the nebula is rotated at, which is shown to be $\approx56$\degr. The 60\degr \, results are being shown as it is close to the estimated angle of rotation. Therefore, the H$\alpha$ and infrared maps suggest that the star is moving at $20 \, \kms$into an ISM with $n\sim100 \, {\rm cm}^{-3}$, with an angle of 60\degr \, with respect to the line of sight.

The apparent spherical emission surrounding the stellar wind bubble in each wavelength is consistent with the observations. Also the brightest part of both emission maps is at the apex of the bow shock which is again consistent with the observations. The `wings' of the bow shock in the H$\alpha$ maps, however, are different to the \textit{HST} image. This part of the bow shock is less bright and fainter in the observational image. The thickness of the predicted bright emission at the rim of the bubble seems consistent with observations near the apex of the bow shock, but at 90\degr \ from the apex we predict a thicker bubble rim than is observed. There are a number of possible reasons for this: a density gradient in the ambient ISM seems to be present \citep{MooWalHesEA02} but is not modelled here; and ISM magnetic fields could alter the thickness of shocked layers.

\section{Calculating X-ray emission}
\label{sec:xray}
\subsection{Synthetic images}
The Bubble Nebula was not detected in X-rays by \textit{ROSAT} \citep{ChuGruGue03} or since then with any other observations. Now we construct synthetic maps of the soft and hard X-ray emission from this nebula and estimate its X-ray luminosity to generate predictions for what X-ray satellites (e.g. \textit{XMM-Newton}) could potentially observe at specific energies. The same raytracing method (described in Appendix B) used to generate the H$\alpha$ emission maps was used to generate the soft and hard X-ray emission maps. The emissivity as a function of temperature for different X-ray bands was calculated using \textsc{xspec} v12.9.1 \citep{1996ASPC..101...17A} and tabulated. Solar abundances from \citet{2009ARA&A..47..481A} as implemented in \textsc{xspec} are used. Absorption within the simulation was neglected but we do consider the effect of interstellar absorption from foreground matter.

\begin{figure*}[htp]
	\centering
	\includegraphics[width=.46\textwidth]{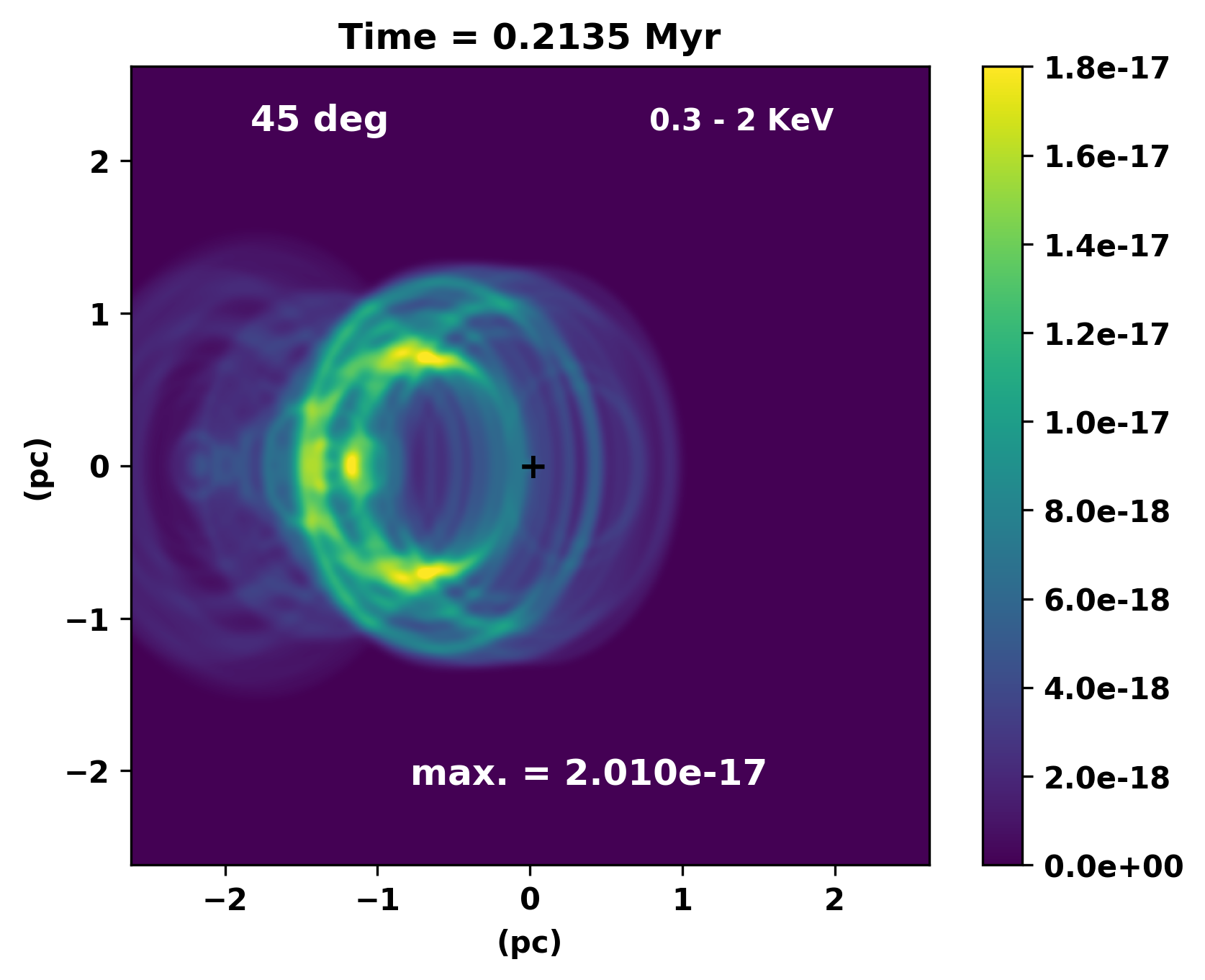}
	\includegraphics[width=.46\textwidth]{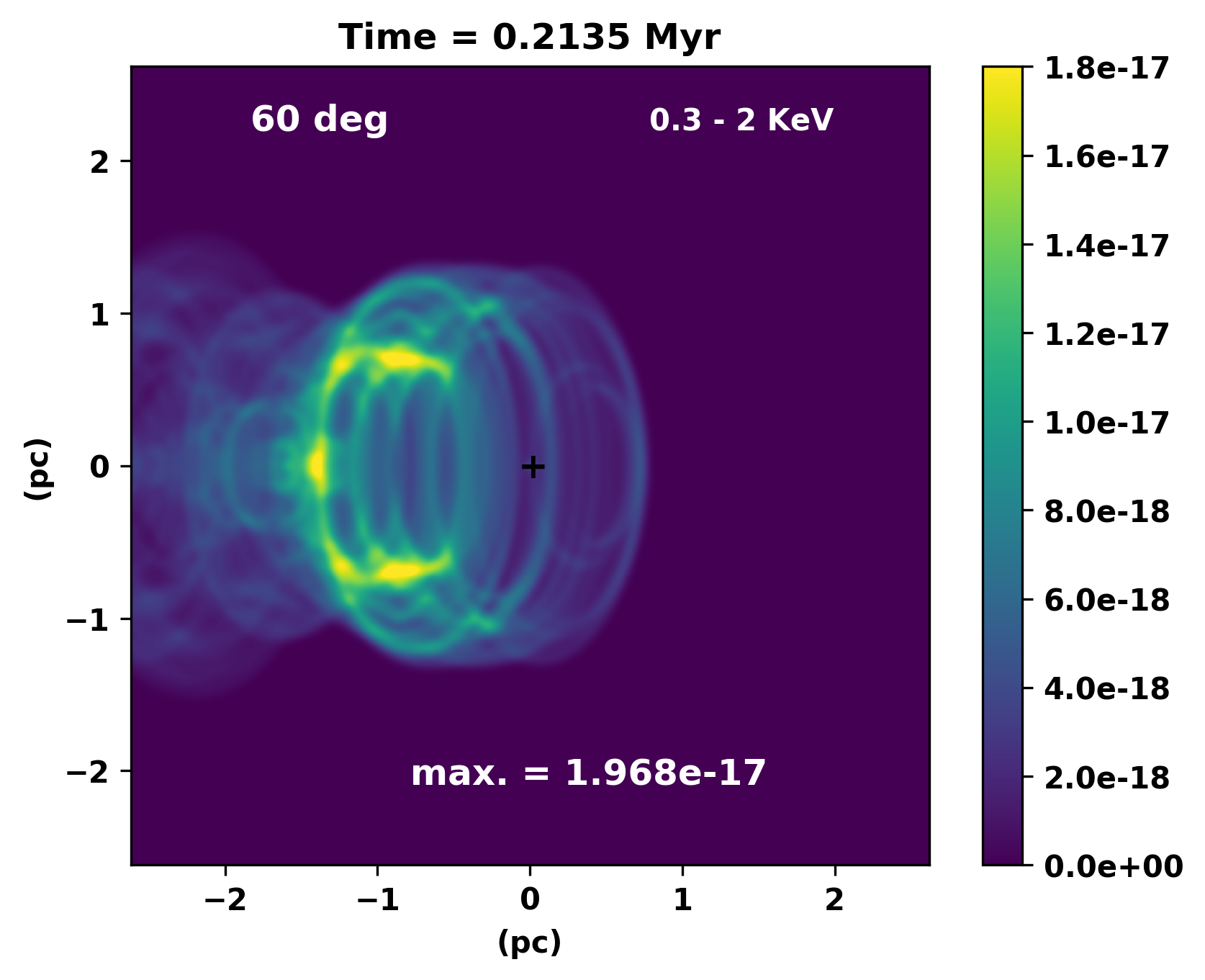}
	
	\caption{Synthetic soft X-ray (0.3--2\,keV) emission maps of the simulated nebula (unabsorbed). Both images have coordinates in pc relative to the star's position (black cross), and scale in $\mathrm{erg\ cm^{-2}\ s^{-1}\ arcsec^{-2}}$. Both images are generated from the 1b simulation, after 0.2135 Myr of evolution and are smoothed to the angular resolution of \textit{XMM-Newton} EPIC cameras (6 arcsec).}
	\label{soft_xray}
	
\end{figure*}

\begin{figure*}[htp]
	\centering
	\includegraphics[width=.46\textwidth]{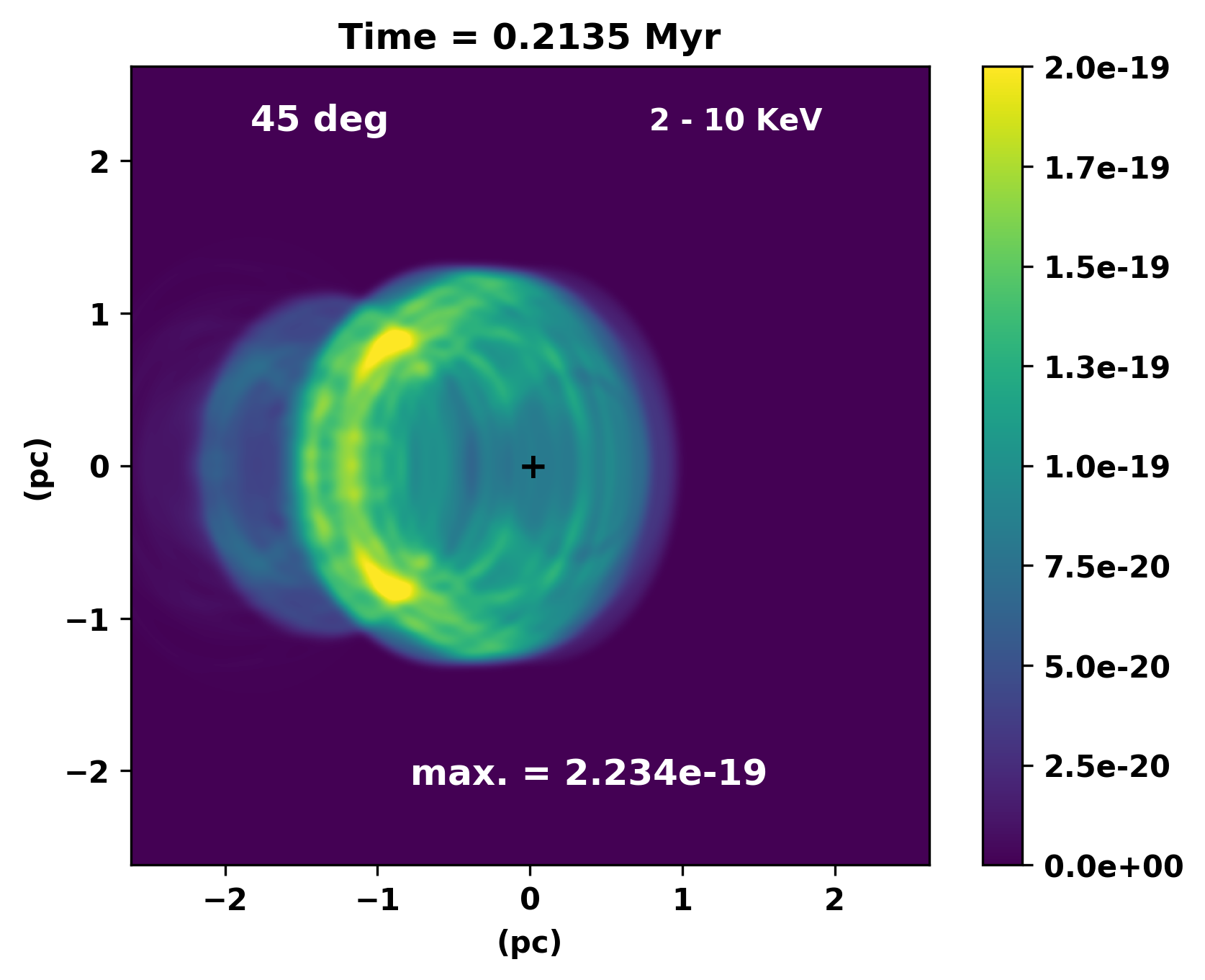}
	\includegraphics[width=.46\textwidth]{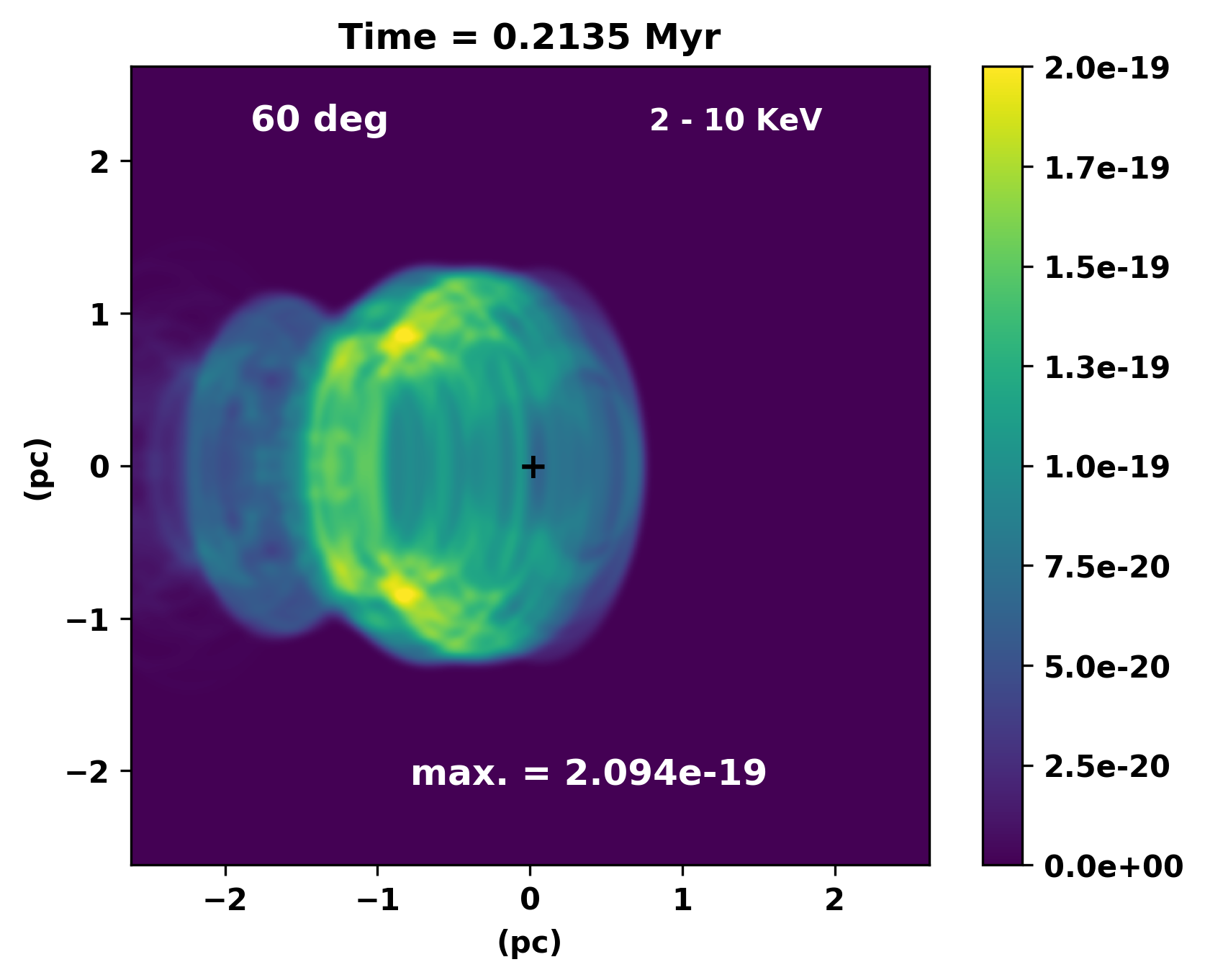}
	
	\caption{As Fig.~\ref{soft_xray}, but for hard X-rays (2--10\,keV).}
	\label{hard_xray}
	
\end{figure*}

\begin{figure*}
	\centering-
	\includegraphics[width=.46\textwidth]{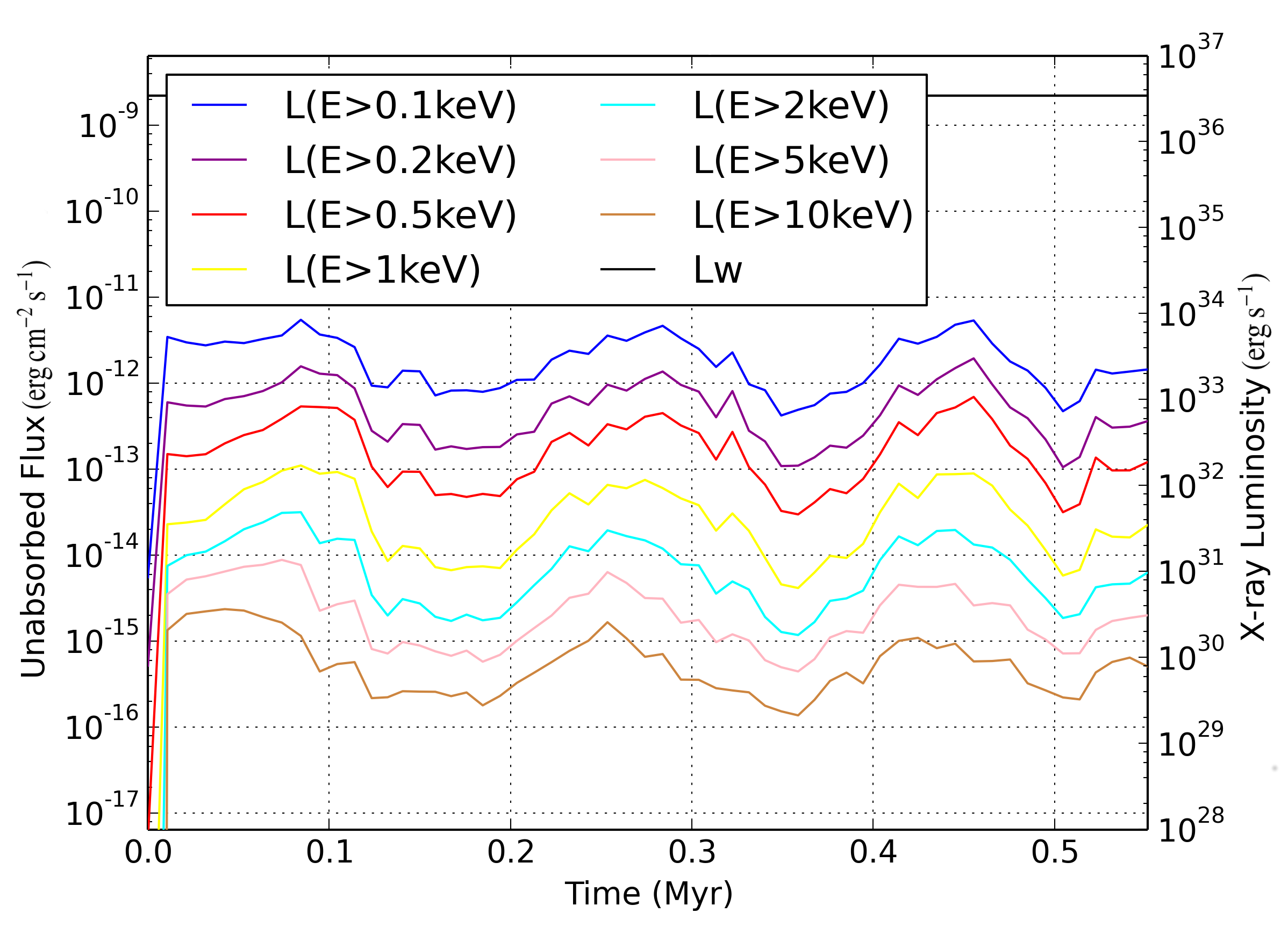}
	\includegraphics[width=.46\textwidth]{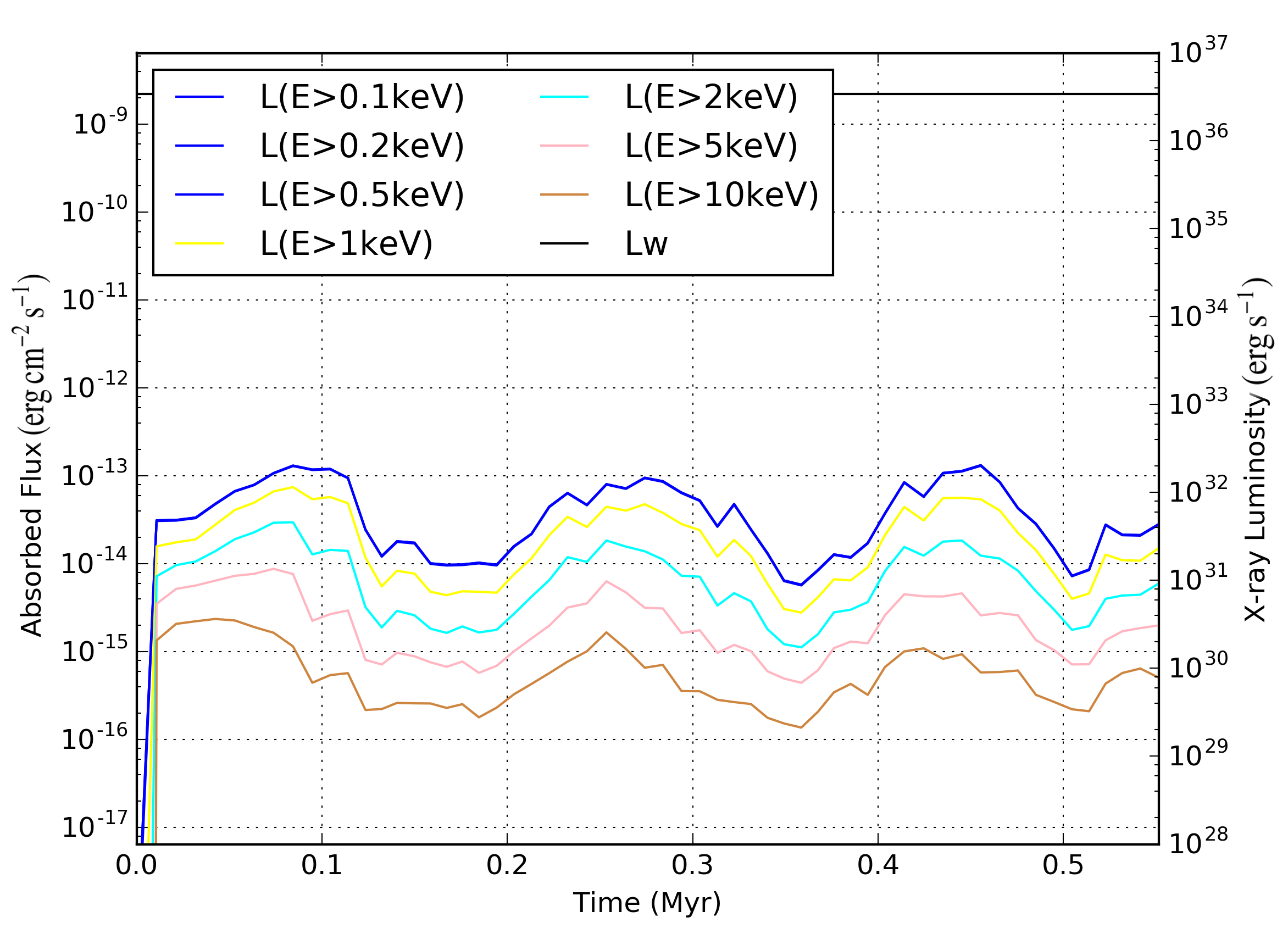}
	
	\caption{Left-hand panel: Synthetic X-ray unabsorbed flux (erg\,cm$^{-2}$\,s$^{-1}$) and luminosity (erg\,$^{-1}$) plot of the Bubble Nebula as it evolves in time (Myr). Seven X-ray bands simulated from soft to hard X-rays, including 0.1\,keV, 0.2\,keV, 0.5\,keV, 1\,keV, 2\,keV, 5\,keV, and 10\,keV. Right-hand panel: Absorbed flux (erg\,cm$^{-2}$\,s$^{-1}$) and luminosity (erg\,s$^{-1}$) plot showing that the 0.1\,keV -- 1\,keV X-ray bands get absorbed by the ISM. Mechanical luminosity of the stellar wind (i.e. energy input rate to the wind bubble) is $L_{\rm w}=\dot{M}v_\infty^2/2 =3.6\times10^{36} \, {\rm erg} \, {\rm s}^{-1}$ (seen as the black line in both plots).}
	\label{lum}
	
\end{figure*}

We generated synthetic soft ($0.3-2$\,keV) and hard ($2-10$\,keV) X-ray emission maps at angles from 0\degr \, to 90\degr \, (with 15\degr \, steps) to the direction of motion for the same snapshot of model 1b as considered in Sections 4 and 5. Fig.~\ref{soft_xray} shows the soft X-ray emission from simulations at angles of 45\degr \, and 60\degr, and Fig.~\ref{hard_xray} shows the same for hard X-ray emission. The soft X-ray band shows plasma around the star (and inside the bubble) hot enough to produce soft X-rays with energies between $0.3-2$\,keV (i.e. temperatures about $10^6$\,K). Whereas, the hard X-ray band shows plasma around the star (and inside the bubble) hot enough to produce hard X-rays with energies between $2-10$\,keV (i.e. temperatures $>10^7$\,K). Both Figs. \ref{soft_xray} and \ref{hard_xray} show that the majority of the X-ray emission is in the `tail' of the stellar wind bubble, as observed for $\zeta$ Oph by \citet{2016ApJ...821...79T}.

Soft X-rays are about 50 times brighter than the hard X-rays with very weak dependence on orientation. Soft X-rays are mostly emitted at the edges of the bubble and are brightest in the wake behind the star. Hard X-ray emission is from the whole volume of the bubble but is also brightest in the wake behind the star. 

The visual extinction towards BD+60\degr2522 of $A_V=2$ mag (Sect.\,\ref{sec:obs}) means that at 0.5 keV only 0.006 of the radiation from the source will reach Earth. At 1 keV the fraction is 0.40, and at 2 keV it is 0.85. This means that $>99$\% of the soft X-rays with $<0.5$\,keV will be absorbed, unless the extinction is patchy or variable across the nebula. Even with this extinction some bright spectral lines could still be observable \citep{2016MNRAS.463.4438T}. The hard X-rays are almost unaffected by extinction, and X-rays with $0.5-2$\,keV are moderately attenuated.

\subsection{Total luminosity}

Fig.~\ref{lum} demonstrates this with the predicted soft and hard X-ray luminosity of the whole nebula. The left image of Fig.~\ref{lum} is a plot of the unabsorbed flux (erg\,cm$^{-2}$\,s$^{-1}$) and the luminosity (erg\,s$^{-1}$) versus the time (Myr) for the whole 1b simulation. The point at each time-step in the graph is the summation of the luminosity/flux over the whole simulated nebula. Each line represents the X-ray luminosity/flux emitted above a certain energy level ranging from 0.1\,keV to 10\,keV. The right plot shows the X-ray flux that could be detected from Earth with extinction taken into account. X-rays with energies $E<0.5$\,keV will get absorbed by the ISM, whereas hard X-rays ($2-10$\,keV) are unaffected. The observable X-ray flux corresponds to a luminosity more than $10^4$ times less than the mechanical luminosity of the stellar wind. The simulations predict a significant X-ray flux in $0.5-2$\,keV which could potentially be observed. The total flux received varies from $10^{-14}$ to $10^{-13}$ erg\,cm$^{-2}$\,s$^{-1}$ depending on time.\footnote{This prediction may be larger than the upper limits derived from unpublished \textit{XMM-Newton} observations (J.Toala, private communication).}

The `dips' in the luminosity seen at $\sim$0.16\,Myr, $\sim$0.35\,Myr, and $\sim$0.5\,Myr coincide with minima in the size of the simulated bubble. For example, the middle image in Fig.~\ref{den} shows a snapshot of the bubble during one of these `dips' in luminosity. When compared to the left image in the same figure, the size of the bubble in the radial direction is very different. Hence, the larger the bubble, the higher the X-ray luminosity emitted. An explanation for why the bubble shrinks in size is that whenever a vortex forms due to the Kelvin-Helmholtz instability at the boundary layer between the bubble and the ISM, it seems to cause a large part of the gas to flow backwards in the $-\hat{z}$ direction. This is known as vortex shedding \citep{2007ApJ...660L.129W}. As the vortex travels backwards in the $-\hat{z}$ direction, it brings all the gas above it with it. The plot in Fig.~\ref{lum} also shows that this is a periodic event. 

\begin{figure}[h]
	\centering
	\includegraphics[width=.49\textwidth]{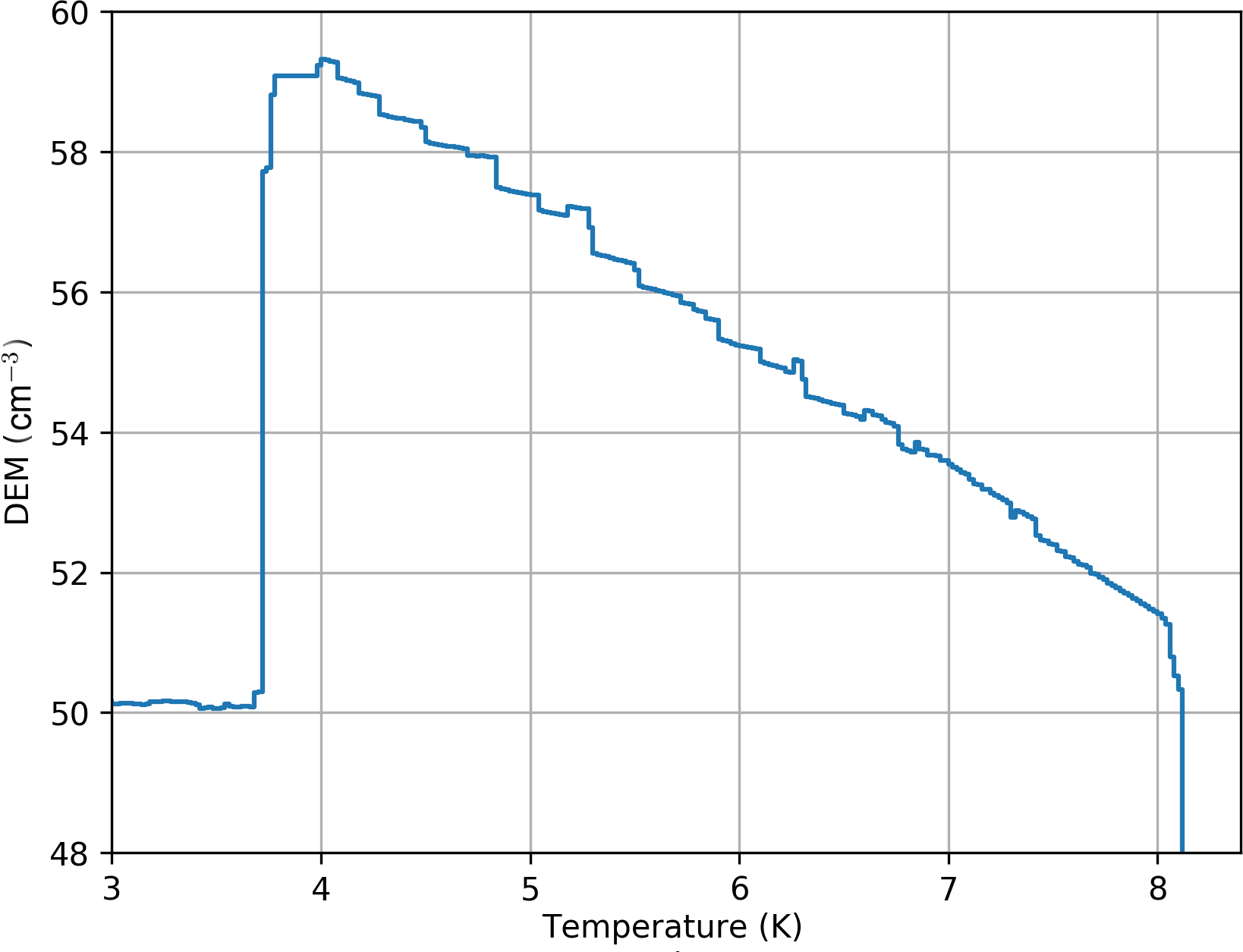} \\
	
	\caption{DEM profile of the simulated nebula (unabsorbed) from the 1b simulation, after 0.2135 Myr of evolution.}
	\label{dem}
	
\end{figure}

\begin{figure}[h]
	\centering
	\includegraphics[width=.49\textwidth]{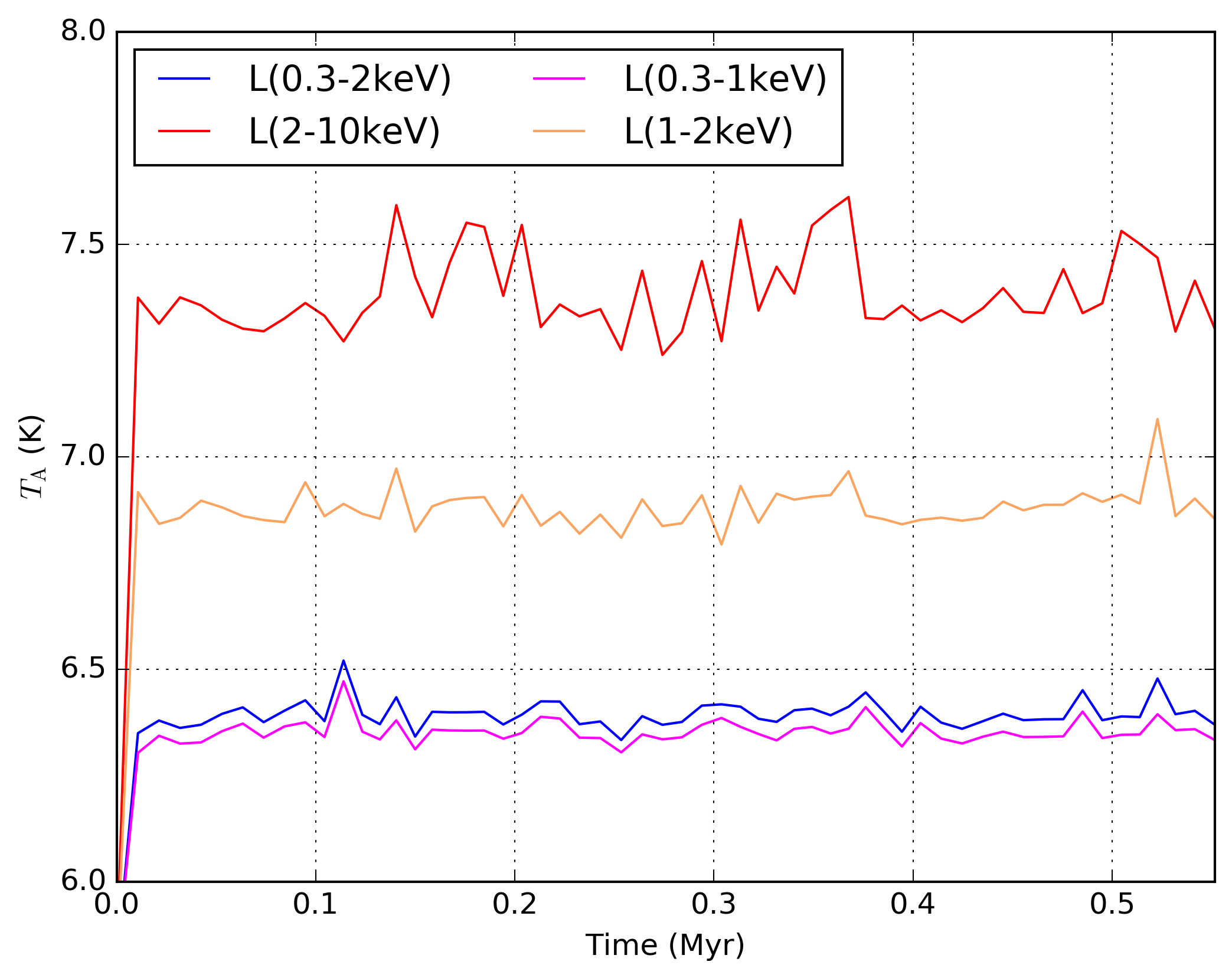} \\
	
	\caption{Mean temperature (K) of the simulated nebula (unabsorbed) as it evolves in time (Myr)}
	\label{dom}
	
\end{figure}

\subsection{Differential emission measure}
The X-ray emission from the hot gas in the 1b simulation is further analysed by calculating the differential emission measure (DEM) as a function of $T$, and then the emission-weighted mean temperature, $T_\mathrm{A}$, following \citet{2018MNRAS.478.1218T}. The DEM profile was calculated using

\begin{equation}
\text{DEM}(T_\mathrm{b}) = \sum_{k, T_\mathrm{k}\in{T_\mathrm{b}}} n_\mathrm{e}^2\Delta{V_\mathrm{k}},
\end{equation}

where $n_\mathrm{e}$ is the electron number density in cell $k$ and $\Delta{V_\mathrm{k}}$ is the volume of cell $k$. All the cells with a gas temperature that fall inside a single bin with median temperature $T_\mathrm{b}$ are summed together. Fig. \ref{dem} shows the DEM profile of the simulated nebula (unabsorbed) from the 1b simulation, after 0.2135 Myr of evolution. The DEM shows a profile strongly skewed towards lower temperatures, with a power-law behaviour similar to that shown by \citet{2018MNRAS.478.1218T} for stellar wind bubbles with turbulent mixing layers and a power-law exponent of $\approx -2$.

We can use the X-ray emissivity in a given energy band, $\epsilon$, together with the DEM profile, to calculate $T_\mathrm{A}$ for the simulated wind bubble, defined by

\begin{equation}
T_\mathrm{A} = \frac{\int \epsilon(T)\text{DEM}(T)T dT}{\int \epsilon(T)\text{DEM}(T) dT}.
\end{equation}

$\epsilon(T)$ is the emission coefficient in the X-ray band and DEM($T$) is the DEM at temperature $T$. The integral is then performed over all the temperature bins in the DEM. Fig. 13 shows the evolution of $T_\mathrm{A}$ as function of time for different X-ray energy bands. The soft X-ray emission shown in Fig. \ref{soft_xray} has a mean temperature of about $10^{6.4}$\,K. Whereas, the hard X-ray emission shown in Fig. \ref{hard_xray} has a mean temperature of about $10^{7.4}$\,K. The figure also shows that the X-ray emission between  $1-2$\,keV has a mean temperature of $10^{6.8}$\,K. The values of $T_\mathrm{A}$ are almost constant for the duration of the simulation, apart from the initial expansion phase of the bubble in the first 0.02 Myr.

\section{Discussion}
\label{sec:discusion}
\subsection{Comparison with the Bubble Nebula}

The original picture that the Bubble Nebula is a supersonically expanding wind bubble has the difficulty that the dynamical age of the nebula of $5\times10^4$\,yr is too young for its associated (moderately evolved) star BD+60\degr2522. For a star moving supersonically through the ISM, however, such an issue does not arise because the timescale for the nebula (bow shock) to reach a stationary state is the maximum of the expansion time of the wind ($R_{\rm SO}/v_\infty$) and the advection timescale of the flow ($R_{\rm SO}/v_\ast$). The latter is much longer for hot stars (such as BD+60\degr2522), but still is $<10^5$ yr for all feasible values of $v_\ast$. The bow shock scenario, therefore, provides an attractive and natural explanation for the apparent youth of the Bubble Nebula. Below we discuss some issues related to this scenario.

The Bubble Nebula is about 3 arcmin in diameter in both the {\it HST} and {\it Spitzer} images. Using the {\it Gaia} distance to BD+60\degr2522 of $d=2.7\pm0.2$ kpc (see Section\,2), this corresponds to a linear diameter of $2.3\pm0.2$ pc. Simulation 1b produces a nebula of diameter 3 pc in both optical and infrared emission (Figs.\,\ref{halpha} and \ref{spitzer}). The model nebula from the 1b simulation is therefore somewhat bigger compared with the observational data from {\it HST} and {\it Spitzer}, implying that some of the input parameters are incorrect. 

Equation (\ref{standoff}) shows that either $\dot{M}$ or $v_\infty$ is too big, or $\rho_\text{ISM}$ or $v_\ast$ is too small. The stellar peculiar transverse velocity of $\approx28 \, \kms$ together with the inclination angle of the bow shock of 60\degr \, imply the total relative velocity of $32 \, \kms$, which is larger than $v_\ast=20 \, \kms$ in our preferred model. A larger $v_\ast$ would prevent the nebula from expanding as much as the 1b simulation. Unfortunately 
our simulations with $v_\ast >30 \, \kms$ became unstable during the simulation runtime because of gas piling up at the apex of the bow shock. This is a well-known problem with 2D hydrodynamic simulations. This problem could possibly be solved by including a magnetic field \citep{2017MNRAS.464.3229M} and/or running more computationally expensive 3D simulations.

\subsection{Mixing/turbulence at the wind/ISM interface}
Only a fraction of the energy input from the massive star's stellar wind goes into the work done to drive the expansion of the bubble. It is thought that the majority of this energy is dissipated by turbulent mixing of the hot shocked wind with the cooler ISM. There is also resultant cooling through line emission or potentially by heat transport through thermal conduction from the wind to the ISM \citep{2014MNRAS.442.2701R}. The turbulent mixing is largely driven by Kelvin-Helmholtz instabilities at the contact discontinuity. 

The soft X-ray emission is coming from the mixing region between the shocked wind and the ISM. Magnetic fields and thermal conduction are not included in these simulations and therefore the contact discontinuity structure is resolved by numerical diffusion and not by physical processes. \citet{2014meyer, 2017MNRAS.464.3229M} showed that both of these processes have some effect, but also that they can cancel each other out somewhat. We will investigate this in future work. The inclusion of a magnetic field can weaken Kelvin-Helmholtz instabilities \citep{1999JPlPh..61....1K,1996ApJ...460..777F}, reducing the amount of mixing which would in turn reduce the intensity of the soft X-ray emission.

\subsection{Limitations of the model}
In our simulations, we have not considered the effect of an ISM density gradient on the structure and appearance of our model wind bubble. The observed H$\alpha$ emission around the Bubble Nebula appears brighter towards the north and gets fainter to the south (see Fig.\,1), indicating that there is a density gradient across the nebula. This is also expected on physical grounds because BD+60\degr2522 photoionizes the surrounding ISM and photoevaporates the molecular cloud to the north. The dense photoevaporated gas expands into the lower density surroundings, creating a density gradient.
	
Arthur \& Hoare (2006) studied wind bubbles expanding into a stratified medium, both with and without stellar motion. In particular, their models H and I considered a star with a strong wind moving through a stratified medium, and they showed that the density gradient induces higher-velocity flows (up to $30 \, \kms$) around the wind bubble from the apex to the tail than are obtained from constant-density calculations. The density gradient that is present in the \hii{} region around the Bubble Nebula could be responsible for the discrepancy between our synthetic (H$\alpha$ and IR) emission maps and observations at the sides of the bubble around 90\degr \, from the apex. In future, more detailed calculations we will explore this in more detail.

\subsection{Importance of winds for particle acceleration and non-thermal processes}
The winds of massive stars generate fast shocks that can accelerate cosmic rays (CR), possibly making a significant contribution to the total high-energy CR production rate in our Galaxy \citep{1980ApJ...237..236C, 1983SSRv...36..173C, 2018arXiv180402331A}.
It is important to identify systems where this can be tested, and where non-thermal emission from relativistic particles can most easily be detected.
The ideal systems have large mass-loss rates with high-velocity winds, but compact nebula surrounded by a relatively dense ISM with a strong magnetic field.
The wind properties maximise the number of accelerated particles, and the ISM properties maximise the interaction of relativistic particles with matter, producing non-thermal radiation.
Searches for non-thermal emission have for this reason concentrated on runaway stars producing bright bow shocks, where the system's geometry is well constrained.
So far the only detection is non-thermal radio emission from the bow shock of the star BD+43\degr3654 \citep{2010A&A...517L..10B}, an O4 supergiant \citep{2007A&A...467L..23C} whose wind drives a large and well-studied bow shock.
Searches for gamma rays \citep{2018A&A...612A..12H} and non-thermal X-rays \citep{2017ApJ...838L..19T} have so far produced only upper limits.
We consider that NGC\,7635 could be a good target for non-thermal emission given the driving star's large mass-loss rate and wind velocity, together with the dense surrounding ISM. BD+60\degr2522 is not as extreme as BD+43\degr3654 in terms of wind mass-loss rate, but their wind velocities are comparable, and the size and density of their nebulae are similar. This suggests BD+60\degr2522 is a good target for further radio observations. These would test whether BD+43\degr3654 is somehow unique or if many bow shocks produce measurable synchrotron emission.

\subsection{X-ray emission resolution study}

\begin{figure}
	\centering
	\includegraphics[width=.49\textwidth]{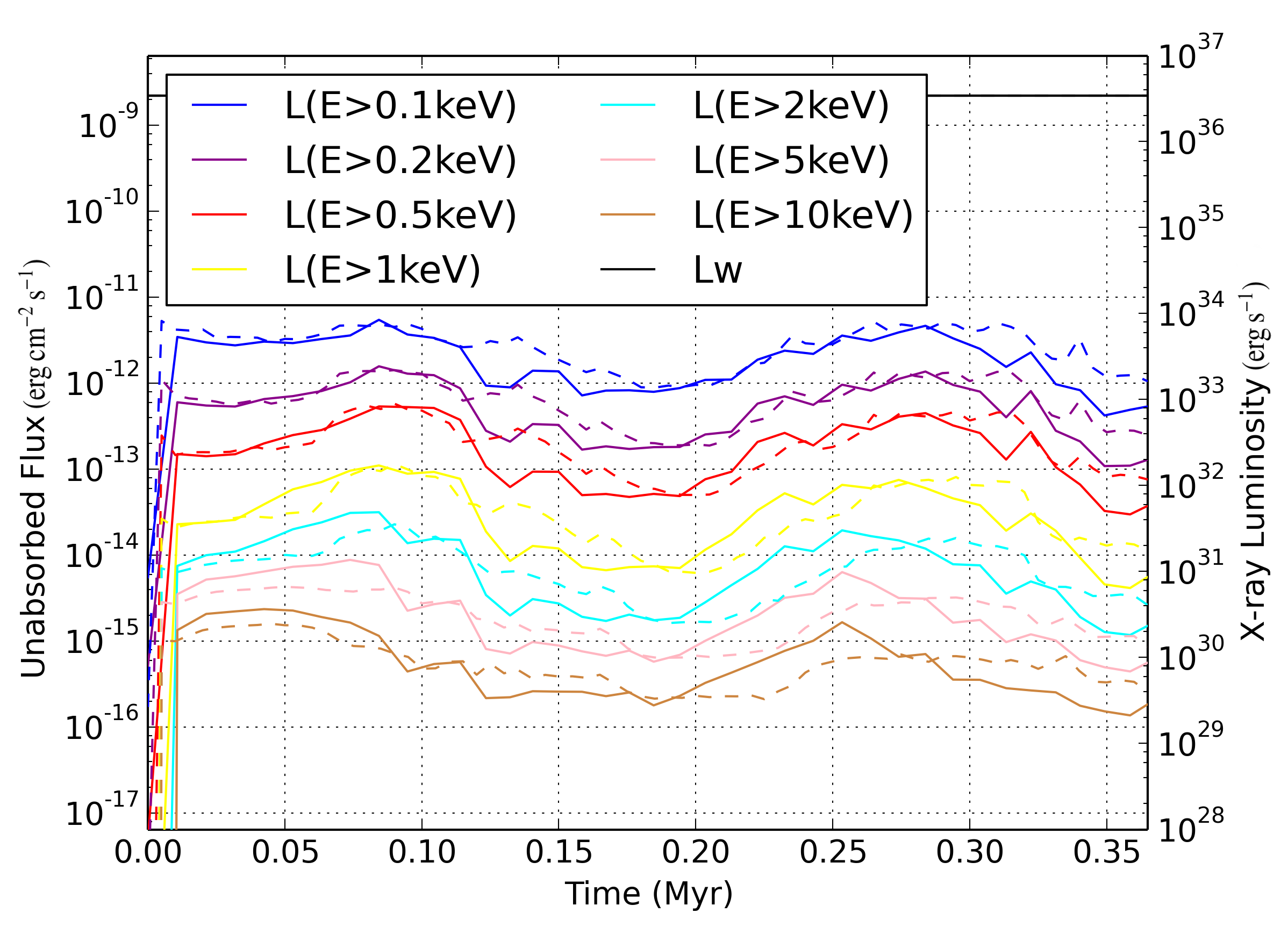} \\

	\caption{Synthetic X-ray unabsorbed flux (erg\,cm$^{-2}$\,s$^{-1}$) and luminosity (erg\,s$^{-1}$) plot of the Bubble Nebula as it evolves in time (Myr). Seven X-ray bands simulated from soft to hard X-rays, including 0.1\,keV, 0.2\,keV, 0.5\,keV, 1\,keV, 2\,keV, 5\,keV, and 10\,keV. Solid lines are from the 1b simulation discussed in this paper, dashed lines are from a lower-resolution ($N_\text{zones}=768\times 512$) simulation. Mechanical luminosity of the stellar wind (i.e. energy input rate to the wind bubble) is $L_\text{w}=3.6\times10^{36}$\,erg\,s$^{-1}$ (seen as the black line in both plots).}
	\label{lum_res}
	
\end{figure}

Fig.~\ref{lum_res} is a plot of the unabsorbed flux (erg\,cm$^{-2}$\,s$^{-1}$) and luminosity (erg\,s$^{-1}$) of the simulated nebula as it evolves in time (Myr) for the whole 1b simulation (solid lines) and for a lower-resolution ($N_{\text{zones}}$ = 768 $\times$ 512) simulation (dashed lines) with the same velocity and ISM density as 1b. This plot compares the two different resolution simulations to show that the results are only weakly dependent on spatial resolution.

\section{Conclusions}
\label{sec:conclusions}
In this paper, we have started a project to investigate thermal emission from stellar wind bubbles. For the first time we present simultaneous predictions for X-ray, optical, and infrared emission maps from simulations. 2D hydrodynamic simulations of the stellar wind bubble NGC\,7635 (Bubble Nebula) have been run to model the interaction of the central star's wind with the ISM. We chose stellar and ISM parameters appropriate for comparison with the Bubble Nebula. The models cover a range of possible ISM densities of $50-200$ cm$^{-3}$ and stellar velocities of $20-40 \, \kms$. One calculation (1b) was found to be the most plausible candidate to compare with observational data.

The Monte-Carlo radiative-transfer code \textsc{TORUS} was used to post-process this simulation to generate synthetic 24$\mu$m emission map predictions to compare with observational \textit{Spitzer} MIPS data. We also post-process the simulation with a raytracing projection code to generate synthetic H$\alpha$ emission maps to compare with \textit{HST} observational data. The main result is that we find the same morphological spherical bubble shape with similar quantitative aspects. The synthetic maps predict a maximum brightness similar to that from the observations and agree that the maximum brightness is at the apex of the bow shock. The H$\alpha$ and 24$\mu$m synthetic emission maps are therefore quantitatively and qualitatively consistent with the observational data and bow shock interpretation. They therefore suggest that the star is moving at $20 \, \kms$ into an ISM with $n\sim100 \, {\rm cm}^{-3}$, at an angle of 60\degr \, with respect to the line of sight.

The raytracing projection code was also used to produce soft ($0.3-2$ keV) and hard ($2-10$ keV) X-ray emission map predictions of what an X-ray satellite could observe. These emission maps show that the majority of X-ray emission occurs in the wake behind the star and not with the bow shock itself. The unabsorbed soft X-rays are in the region of $\sim 10^{32} - 10^{33}$ erg\,s$^{-1}$. However, due to extinction from the ISM in between the nebula and the observer, no X-rays below $0.5$ keV can be seen and X-rays between ($0.5-2$\,keV) are significantly attenuated. The hard X-rays are faint, $\sim 10^{30} - 10^{31}$ ergs\,s$^{-1}$, and maybe too faint for current X-ray instruments to successfully observe.

Results from the simulations and the synthetic emission maps allow us to conclude that the O star creates a bow shock as it moves through the ISM and in turn creates an asymmetric bubble visible in optical and infrared wavelengths, and predicted to be visible in X-rays. The Bubble Nebula does not appear to be unique, it just has a favourably oriented very dense bow shock. Extinction means UV and soft X-rays will be hard to detect and therefore it is hard to constrain the mixing between the hot and cold plasma. However, the dense ISM surrounding BD+60\degr2522 and its strong wind makes it a good candidate for detecting non-thermal emission at other wavelengths.
\begin{acknowledgements}
SG is funded by a Dublin Institute for Advanced Studies student scholarship. We acknowledge the SFI/HEA Irish Centre for High-End Computing (ICHEC) for the provision of computational facilities and support to run the {\sc pion} simulations of the Bubble Nebula (project dsast018b). JM acknowledges funding from a Royal Society-Science Foundation Ireland University Research Fellowship (14/RS-URF/3219). TJH is funded by an Imperial College Junior Research Fellowship. VVG acknowledges support from the Russian Science Foundation grant No. 14-12-01096. SG wishes to thank Associate Professor Ben Thornber, from the School of Aerospace, Mechanical and Mechatronic Engineering at the Univeristy of Syndey, for very useful discussions on the concept of vortex shedding. SG and JM also wish to thank Norbert Langer for hosting discussion visits to Bonn. The authors are grateful to the referee, J.~Toal\'a, for constructive suggestions that improved the manuscript. The Darwin Data Analytic system at the University of Cambridge, operated by the University of Cambridge High Performance Computing Service on behalf of the STFC DiRAC HPC Facility (www.dirac.ac.uk) was used to run \textsc{TORUS} to simulate the emission maps. This equipment was funded by a BIS National E-infrastructure capital grant (ST/K001590/1), STFC capital grants ST/H008861/1 and ST/H00887X/1, and DiRAC Operations grant ST/K00333X/1. DiRAC is part of the National E-Infrastructure. Some of the data presented in this paper were obtained from the Mikulski Archive for Space Telescopes (MAST). 
STScI is operated by the Association of Universities for Research in Astronomy, Inc., under NASA contract NAS5-26555.
This work has made use of the NASA/IPAC Infrared Science Archive, which is operated by the Jet Propulsion Laboratory,
California Institute of Technology, under contract with the National Aeronautics and Space Administration, and the 
SIMBAD database, operated at CDS, Strasbourg, France. This work also has made use of 
data from the European Space Agency (ESA) mission {\it Gaia} (\url{https://www.cosmos.esa.int/gaia}), processed by the 
{\it Gaia} Data Processing and Analysis Consortium (DPAC, \url{https://www.cosmos.esa.int/web/gaia/dpac/consortium}). 
Funding for the DPAC has been provided by national institutions, in particular the institutions participating in the 
{\it Gaia} Multilateral Agreement.
\end{acknowledgements}

\bibliographystyle{aa}
\bibliography{./refs}

\appendix
\section{Peculiar transverse velocity of BD+60\degr2522}
\label{sec:appendixA}

\begin{table*}[t]
	\caption{Summary of astrometric and kinematic data on BD+60\degr2522.}
	\label{tab:prop}
	\centering
	\begin{tabular}{cccccc}
		\hline 
		$d$ & $\mu _\alpha \cos \delta$ & $\mu _\delta$ & $v_{\rm l}$ & $v_{\rm b}$ & $v_{\rm tr}$ \\
		(kpc) & (mas ${\rm yr}^{-1}$) & (mas ${\rm yr}^{-1}$) & ($\kms$) & ($\kms$) & ($\kms$) \\
		\hline 
		$2.7\pm0.2$ & $-2.71\pm0.05$ & $0.53\pm0.05$ & $11\pm2$ & $25\pm2$ & $28\pm3$ \\ 
		\hline
	\end{tabular}
\end{table*}

{\it Gaia} DR2 \citep{2018A&A...616A...1G} places BD+60\degr2522 at the distance of $2.7\pm0.2$ kpc and provides high-precise proper motion measurements for this star (see Table\,\ref{tab:prop}). Using the solar galactocentric distance of 8.0\,kpc, the circular Galactic rotation velocity of $240 \, \kms$ \citep{2009ApJ...705.1548R}, and the solar peculiar motion $(U_{\odot},V_{\odot},W_{\odot})=(11.1,12.2,7.3) \, \kms$ \citep{2010MNRAS.403.1829S}, we calculated 
the peculiar transverse velocity $v_{\rm tr}=(v_{\rm l}^2 +v_{\rm b}^2)^{1/2}$, where $v_{\rm l}$ and $v_{\rm b}$ are the star's peculiar velocity components along the Galactic longitude and latitude, respectively. For the error calculation, both the uncertainties in the proper motion and the distance measurements were considered. The resulting velocities along with the input data are given in Table\,\ref{tab:prop}.

\section{Raytracing of 2D simulations}
\label{sec:appendixB}

Here we describe a raytracing method to calculate synthetic images from such simulations. The simplest approach 
is to take the 2D grid of zones, consider a ray going through each zone-centre in turn, and then produce an image 
with the same number of pixels as the grid has zones. The grid is then extended in the $z$-direction to include all 
rays that intersect some part of the 3D volume created by rotating the 2D plane about the axis of symmetry.

Consider a ray travelling through Cartesian space $(x,y,z)$, in the plane $y=R_0$ (where $R_0$ is the distance from the axis of symmetry), with an angle $\theta$ with 
respect to the positive $z$-axis, and with the equation $x=(z-z_0)\tan\theta$. We can place the $R-z$ plane 
as the upper-half plane $x=0, y\geq0$. This can be done without loss of generality such that the ray passes 
through $(x,y,z)=(0,R_0,z_0)$, corresponding to a grid zone with coordinates ($R_0, z_0$). 

We now calculate the path of the ray when projected onto the $R-z$ plane. An infinitesimal line element is 
$\vec{d\ell}=\cos\theta\vec{dz} + \sin\theta\vec{dx}$ and so
\begin{equation}
dx = \pm \frac{RdR}{\sqrt{R^2-R_0^2}}.
\end{equation}
Using the equation of the line above, $dx=dz\tan\theta$, we can get 
\begin{equation}
dz = \pm \frac{RdR}{\tan\theta\sqrt{R^2-R_0^2}}.
\end{equation}
Therefore, the geometric scaling factor is 
\begin{equation}
d\ell = \frac{RdR}{\sin\theta\sqrt{R^2-R_0^2}}.
\end{equation}
Furthermore, it is easy to show that the ray traces a parabola in the $R-z$ plane:
\begin{equation}
z = \pm \frac{\sqrt{R^2-R_0^2}}{\tan\theta} + z_0, \, R = \sqrt{R_0^2 + (z-z_0)^2\tan^2\theta}.
\end{equation}
If we approximate our data as a piecewise-constant, with each zone having constant values of each variable, then we can 
analytically integrate the emissivity along a ray segment through a zone, $i$, as
\begin{align}
\int j(\vec{x})\vec{d\ell} = \int_{R_-}^{R_+} j(R_i, z_i) \frac{RdR}{\sin\theta\sqrt{R^2-R_0^2}} \nonumber \\
= \frac{j(R_\mathrm{i}, z_\mathrm{i})}{\sin\theta}\left(\sqrt{R_+^2 - R_0^2} - \sqrt{R_-^2 - R_0^2}\right),
\end{align}
where the ray enters the zone at $R = R_-$ and leaves at $R = R_+$. The sign 
ambiguity is resolved by considering that the ray is 
always moving to smaller $R$ on the inward trajectory and larger $R$ on the outward one, but in both cases the emissivity 
adds to the quantity being integrated.

\end{document}